\documentclass{article}[12pt]
\usepackage[T1,T2A]{fontenc}
\usepackage[cp1251]{inputenc}
\usepackage{amsmath,amssymb}
\usepackage{graphicx}%
\usepackage{xcolor}
\usepackage{verbatim}
\usepackage{multirow}

\newtheorem{note}{Note}
\newtheorem{stat}{Statement}

\newcommand{\Lee}[1]{\stackunder{#1}{\rm L}}
\newcommand{\Vivod}[1]{\noindent\hspace{12pt}$\bullet$ \hskip 4pt #1\\[6pt]}
\def\stackunder#1#2{\mathrel{\mathop{#2}\limits_{#1}}}
\begin{document}

\begin{center}
{\bf \Large Yu. G. Ignat'ev\footnote{Yu. G  Ignat'ev, Institute of Physics, Lobachevsky Institute of Mathematics and Mechanics, Kazan Federal University, Kremlyovskaya str., 35, Kazan, 420008, Russia. email: yurii.ignatev.1947@yandex.ru}, D. Yu. Ignatyev\footnote{Institute of Physics, Kazan Federal University, Kremlyovskaya str., 35, Kazan, 420008, Russia; email: ignatyev.id@gmail.com}} \\[12pt]
{\bf \Large Cosmological Models Based on a Statistical System of Scalar Charged Degenerate Fermions and an Asymmetric Higgs Scalar Doublet} \\[12pt]
\end{center}

\abstract{On the basis of the general relativistic statistical and kinetic theory, a consistent closed cosmological model is formulated. It is based on a statistical system of scalar charged fermions interacting by means of classical and phantom scalar fields.
Based on the study of the microscopic dynamics of scalar charged particles, within the framework of the Lagrangian as well as Hamiltonian formalism, a function of the dynamic mass of scalar charged particles was constructed and it was shown that for the consistency of the theory, it is necessary to remove the nonnegativity condition for this function.
On the basis of the Lagrangian formalism, equations of gravitational and scalar fields with singular sources are formulated and microscopic conservation laws are obtained.
Within the framework of the general relativistic kinetic theory, macroscopic equations of gravitational and scalar fields are formulated and macroscopic conservation laws are obtained. These equations' full correspondence to microscopic equations with singular sources is shown.
Further, on the basis of the obtained equations, a cosmological model for a degenerate system of scalarly charged fermions is formulated.
An exact solution of the constitutive equations for a degenerate scalar-charged plasma in the cosmological model is obtained, which made it possible to significantly simplify the original system of equations.
On the basis of the obtained solution of the constitutive equations, two fundamentally different cosmological models are formulated, one of which has two types of singly scalarly charged fermions, while the second has one kind of fermions charged with two charges of various nature. A qualitative analysis of the obtained 6-dimensional dynamic system for a two-component model is carried out. It is shown that in such models, acceleration deceleration modes become possible at the later stages of the evolution of the Universe.  
}

%
\section*{Introduction}
\quad From a formal point of view, phantom fields appear to have been introduced into gravity as one of the possible scalar field models in 1983 in \cite {Ignat83_1}. In this work, as well as in later ones (see, for example, \cite{Ignat_Kuz84}, \cite{Ignat_Mif06}) phantom fields were classified as scalar fields with the attraction of like-charged particles and were identified by the factor $\epsilon = -1$ in the energy tensor - the momentum of the scalar field. Note that phantom fields in relation to wormholes and black universes were considered in the works \cite{Bron1}, \cite{Bron2}.
However, the introduction of phantom fields into the structure of quantum field theory encounters serious problems associated either with the probabilistic interpretation of quantum theory, or with the problem of the stability of the vacuum state, due to the unboundedness of negative energy \cite{Cline}. The negative kinetic term and violation of the isotropic energy condition imply that the energy is not bounded from below at the classical level, so negative norms appear at the quantum level. In turn, negative norms of quantum states generate negative probabilities, which contradict the standard interpretation of quantum field theory \cite{Linde}, \cite{Saridakis}. The requirement that the theory be unitary leads to instability when describing the interaction of quantized phantom fields with other quantized fields \cite{Sbisa}.

In the paper \cite{Vernov}, however, it is noted that the terms leading to instability can be considered as corrections that are significant only at low energies below the physical cutoff. This approach allows us to consider phantom field theories as some effective, physically acceptable theories, while it is assumed that an effective theory allows immersion in some fundamental theory, for example, string field theory, which is consistent with the well-known idea of S.M.  Carroll, M. Hoffman and M. Trodden on effective field theory, in which the phantom model can be viewed as part or sector of the more fundamental \cite{Trodden} theory (see also \cite{Richarte}). In particular, in these works it was shown that there is a low-energy boundary, such that the phantom field will be stable during the lifetime of the Universe.

On the other hand, an analysis of the observations of the cosmological acceleration, as well as the associated \emph{barotropic coefficient} $w= p/\varepsilon$, carried out by various groups of researchers in recent years, shows that, apparently, it will be very difficult to do without phantom fields in cosmology. For example, SNIa data show a significant preference for "phantom" models and exclude the cosmological constant \cite{Tripathi}. Strong restrictions can be obtained in combination with other observational data, including measurements of the Hubble parameter $H(z)$ at different redshifts. When combining standard rulers and standard clocks, the best match is observed when $w_0=-1.01(+0.56 -0.31)$ \cite {Ma}. For the flat wCDM model, the constant parameter of the dark energy equation of state $w=-1.013 (+0.068 -0.073)$ \cite{Meyers} was measured, see also \cite{Terlevich, Chavez}.

In a number of works by I. Ya. Aref'eva, S. Yu. Vernov, AS Koshelev, et al. \cite{Vernov1} -- \cite{Vernov3}, two-field cosmological models based on a pair of scalar fields, classical and phantom, were investigated. in which a negative kinetic term corresponds to the phantom field. In this case, under the assumption of the polynomial potential of scalar fields of the 6th order, the classes of one-parameter and two-parameter exact solutions were found. As noted in \cite{Vernov2}, \cite{fantom} phantom fields are involved in cosmology to provide the value of the \emph{barotropic} coefficient $w<-1$ ($w: \ p = w \varrho$) in order to prevent `` the big break''. Note that in these papers, two-field cosmological models are substantiated by string theory, in which a tachyon describing brane decay is considered as a phantom field.

In the works of Yu.G. Ignatiev, A.A. Agafonov and I.A. Koch carried out a comprehensive study of incomplete cosmological models under the assumption that the Hubble constant is nonnegative for the cases of the classical Higgs vacuum field \cite{YuI_Clas17}, the Higgs phantom field \cite{YuI_Fant17}, \cite{YuI_Agaf_Fant17} and the asymmetric scalar RR doublet (quintom) \cite{YuI_Kokh_Russ18_1} -- \cite{YuI_Kokh19_2}. In such models, transitions of cosmological evolution from the stage of expansion to the stage of contraction become possible (and, conversely, for a phantom field. If we discard a number of incorrect results of works, just related to the assumption of the nonnegativity of the Hubble constant, then one of the results of these works can be summarized as follows:  in the late stages of evolution, the cosmological model, \emph{based on vacuum scalar fields}, always reaches inflation. The same result was confirmed by studies of the full model \cite{Ignat20}, in which the assumption of the non-negativity of the Hubble constant was removed. Finally, in the work \cite{Ignat21_TMP}, a complete mathematical model of the cosmological evolution of the asymmetric vacuum scalar Higgs doublet (quintom) was investigated and it was shown that in this model transitions from the cosmological compression regime to the expansion regime, and vice versa, and also possible oscillatory modes with a change in the phases of compression and expansion. Note that in the article by Yi-Fu Cai, Emmanuel N. Saridakis, Mohammad R. Setare and Jun-Qing Xia \cite{Saridakis2010}, the prospects of quintom cosmology were linked precisely in connection with the possibility of obtaining a cosmological scenario that would allow avoiding the Big Bang singularity. In addition, we note that, contrary to popular belief, the phantom field plays the role of a stabilizer for the stable accelerated expansion of \cite {Ignat21_TMP}, thus being a necessary additional component of the cosmological model. Although the oscillatory regime discovered in this work \cite{Ignat21_TMP} has not been sufficiently investigated, in all other cases vacuum quintom models exhibit the following asymptotic behavior \footnote{Of course, this conclusion does not apply to cosmological models that include, in addition to scalar fields and \ newline, other material components, see for example \cite{Chiba}}:
\begin{equation}\label{H8,O8}
H(\pm\infty)\to\pm |H_0|,\qquad \Omega(\pm\infty)\to 1,
\end{equation}
where $H(t)$ is the expansion rate (Hubble parameter) and $\Omega(t)$ is the invariant cosmological acceleration:
\begin{equation}\label{H=,Omega=}
H=\frac{\dot{a}}{a};\quad \Omega=\frac{\ddot{a}a}{\dot{a}^2},
\end{equation}
$H_0$ is a constant that depends on the parameters of the model.
We note a similar independent work in 2018 by G. Leon, A. Paliathanasis and J. L. Morales \cite{Leon18}, in which a detailed qualitative analysis of the quintom cosmological model with exponential potential energy of the classical and phantom scalar fields was carried out.

On the other hand, in a number of earlier works based on the theory of statistical systems of scalar charged particles, \cite{Ignat14_1} - \cite{Ignat15_1}, in which the cosmological evolution of such systems was investigated, the possibility of four types of behavior of the corresponding cosmological models was shown , among which there were models with an intermediate ultrarelativistic stage and a final nonrelativistic \cite{Ignat15_2} -- \cite{Ignat17}. However, these studies were based, firstly, on an incomplete mathematical model, secondly, on the quadratic potential of scalar fields and, thirdly, on a scalar singlet. In the work \cite{Ignat20_1}, a cosmological model of the cosmological evolution of the statistical system of degenerate scalar charged fermions interacting by means of a single scalar Higgs field, classical or phantom, is formulated. This article also provides examples of numerical models of such systems, which radically differ in their behavior from the behavior of models based on vacuum scalar fields. Finally, in the works of one of the authors \cite{Ignat_Stability_1,Ignat_Stability_2}, a mathematical model of gravitational perturbations of the cosmological system of scalar charged fermions in the case of a scalar Higgs singlet (classical or phantom) was constructed and the stability of this cosmological model with respect to short-wave and scalar perturbations of the gravitational field was investigated. In these works, it was shown that the fermionic system is unstable in the early stages of cosmological expansion in the case of the classical Higgs interaction and is stable in the case of the phantom Higgs interaction.

These and other unique features of statistical systems of scalar charged fermions indicate the need for a more detailed theoretical analysis of such systems, firstly, a rigorous microscopic and macroscopic substantiation of the theoretical model of the interaction of particles with scalar fields and, secondly, revealing the global properties of cosmological models based on statistical systems of scalar charged particles. It should be noted that the construction of a mathematical model for systems of scalar charged particles encounters a number of serious theoretical problems associated, for example, with the determination of the total mass of particles in a scalar field, which require careful analysis. In this regard, the problem arises of formulating a complete mathematical model of cosmological systems of scalar charged particles with Higgs scalar fields, including an asymmetric scalar doublet.

It should be noted that from the late 1990s up to the present time, many researchers (B. Saha, G.N. Shikin, MO Ribas, FP Devecchi, GM Kremer, L. Fabbri, J. Wang, S.-W. Cui , S.-M. Zhang, Yu.P. Rybakov, K.A. Bronnikov, T. Boyadzhiev and others) are studying cosmological models based on scalar and nonlinear fermionic (spinor) fields (see, for example, the articles \cite{Saha97} -- \cite{Bron_Rybak2020} and the literature cited therein). Interest in nonlinear spinor fields in cosmology is due to the fact that they can be used to solve a number of problems in cosmology: 1) accelerate the process of isotropization of an initially anisotropic Universe; 2) ensure the absence of cosmological singularity, 3) generate a late cosmological acceleration. In addition, the nonlinear spinor field is a fairly good model of both an ideal fluid and various types of dark energy. In particular, J. Wang et al. \cite{Wang} investigated the cosmological model based on the spinor quintom.

In our article, we will formulate a mathematical model of the cosmological statistical system of classical scalar charged particles with Higgs scalar fields, investigate its main properties and carry out its qualitative analysis for simple models of the scalar particle charge in the Higgs doublet field. In this article, we will consider a statistical system of scalarly charged degenerate fermions as a plasma model. The interest in this model is due to two circumstances. First, in the case of complete degeneracy, the expressions for macroscopic plasma scalars are calculated in elementary functions, which allows a detailed analysis of such a system. The second, more serious circumstance is as follows. According to the well-known scenario, a degenerate Fermi system with interparticle interaction can transform into a Bose condensate with properties similar to those of superfluidity and superconductivity (scalar) by the formation of Cooper pairs of like scalarly charged fermions. This condensate, as a system with minimum energy, in principle, can be added to the scalar vacuum in cosmological models and realize the observed component of dark bosonic matter. The fact that under conditions of sufficiently strong phantom scalar fields the degree of degeneracy of scalar-charged fermions grows in the process of cosmological expansion \cite{Ignat17}.

\section{The Motion of a Scalar Charged Particle in Scalar and\newline Gravitational Fields }

\subsection{Lagrangian Formalism for the Description of the Motion of a Scalar Charged Particle}
We will refer here to the works \cite{Ignat15} and\cite{Ignat15_2}, which contain correct generalization of the relativistic theory for both the case of phantom scalar fields and the sector of negative dynamic masses of scalar charged particles.
Let us consider the motion of a scalar charged particle with scalar charges $q^r$ in scalar fields described by the potentials $\Phi^r(x^j)$  ($r=\overline{1,N}$). The only way to generalize the action function of a massive free particle in a gravitational field to the case of a scalar charged particle is the following \cite{Ignat15_2} \footnote{Here and below, the Planck system of units is chosen  $G=c=\hbar=1$.}
\begin{equation}\label{S_m0}
S=-\int F(\Phi^1,\ldots,\Phi^N)ds,
\end{equation}
where $F(\Phi^1,\ldots,\Phi^N)$ - is a certain given function of scalar potentials. By varying the action \eqref{S_m0} along the trajectory of the particle, we obtain the equations of motion \cite{Ignat15_2}:
\begin{equation}\label{du/ds}
\frac{\delta u^i}{\delta s}=\partial_k\ln|F|\ \pi^{ik}(u),
\end{equation}
where $u^i(s)=dx^i/ds$ -- is a velocity vector of a particle, $\delta/\delta s$ -- are absolute derivatives along the trajectory $x^i=x^i(s)$ (see, for instance, \cite{Sing}):
\[\frac{\delta u^i}{\delta s}=\frac{du^i}{ds}+\Gamma^i_{kl}u^k\frac{du^l}{ds}\equiv u^k\nabla_k u^i,\]
$\nabla_k$ -- is an operator of covariant differentiation with respect to the metric $g_{ik}$\footnote{In the future, where convenient, we denote the covariant differentiation operator by comma:\\ $\nabla_i T=T_{,i}$, $\nabla^i T=T^{,i}$.}. Further on $\pi^{ik}(u)$ in \eqref{du/ds} -- is a \emph{symmetrical operator of orthogonal projecting on the unit vector} $u^i$:
\begin{equation}\label{pi_ik}
\pi^{ik}(u)= \bigl(u^iu^k-g^{ik}\bigr)
\end{equation}
so that
\begin{equation}\label{pi_iku^k}
\pi_{ik}(u)u^k=0; \quad \pi^{ik}(u)g_{ik}=-n+1=-3.
\end{equation}

The first integral of the equations of motion for any functions $F(\Phi^1,\ldots,\Phi^N)$ is the normalization relation $(u,u)=\mathrm{Const}$. From the comparison with the action function of a free massive particle of mass $m_0$ in the gravitational field (see, for instance, \cite{Land}) it follows
\begin{equation}\label{F(0)=m0}
F(0,\ldots,0)=m_0,
\end{equation}
where $m_0$ -- is a bare mass of a particle in absence of scalar fields.

The fundamental principle of additivity of the Lagrange function taking into account the limit relation \eqref{F(0)=m0} leads us to the only possible form of the action function \eqref{S_m0}, and at the same time, the function $F(\Phi^1,\ldots)$:
\begin{eqnarray}\label{m*}
F\equiv m_*=m_0+\sum\limits_{r=1}^N  q^r\Phi_r;\\
\label{S_m1}
S=-m_0\int ds-\sum\limits_{r=1}^N q^r\int\Phi_r ds\equiv -\int m_*ds,
\end{eqnarray}
where $m_*$ -- is a dynamic rest mass of a particle. Note that in the extended theory \cite{Ignat15} the dynamic mass \eqref{m*} can also be negative. However, it is this form of the dynamic (inert) mass of a particle that is a direct consequence of one of the fundamental principles - the principle of additivity of the Lagrange function. An attempt to enclose the right-hand side of the equality \eqref{m*} under the modulus sign leads to a violation of the additivity principle and, at the same time, to a number of additional problems. Let us note also that the equations of motion of a scalar charged particle \eqref{du/ds} are invariant with respect to substitution $m_*\to -m_*$.

Note that in this notation the equations of motion of the particle \eqref{du/ds} take the form:
\begin{equation}\label{du/ds_m}
\frac{\delta u^i}{\delta s}=\partial_k\ln|m_*|\ \pi^{ik}(u),
\end{equation}

The kinematic momentum of the particle $p^i$ lies on the dynamic mass shell \cite{Ignat15}:
\begin{equation}\label{8_1}
(p,p)=m^2 \Rightarrow \tilde{p}\ \!^4=\sqrt{m^2_*+\tilde{p}^2},
\end{equation}
where $\tilde{p}^{(i)}$ -- are reference projections of the momentum vector, $p^2$ -- is a quadrate of the physical momentum. Note that the dynamic mass of particles is not a heavy mass, which, in contrast to the dynamic mass, is determined by the total energy, i.e., $m=\sqrt{m_*^2+p^2}$, so $m(p=0)=|m_*|$ (see. \cite{Ignat15}). The symmetry requirement between particles and antiparticles ($q\to -q$) leads to the condition that the bare mass is equal to zero in the formula \eqref{m*} \cite{Ignat15}. Accordingly, further on, for the dynamic mass of a particle of the ``$a$'' we assume:
\begin{equation}\label{m_*(pm)}
m_*\equiv  m_{(a)}=\sum\limits_r q_{(a)}^r\Phi_r.
\end{equation}
Thus, we do not exclude the possibility of a negative dynamic mass of fermions, in particular, the dynamic masses of particles and antiparticles will differ in sign in this case while the total rest masses will coincide\footnote{We highlight the index of the type of particles, $a$ with parentheses , in order to avoid confusion}. Hence, the \emph{total rest mass of a particle} $m$ is equal to
\[m=|m_{(a)}|.\]

\subsection{Invariant Canonical Formalism for Describing the Motion of a Scalar Charged P,article }
The previous consideration reveals the problem of correct determination of the mass of a scalar charged particle, which requires a deeper study and without solving which it is impossible to build a correct mathematical model of scalar charged particles. In this regard, let us turn to the invariant canonical (Hamiltonian) formalizm of the motion equations' definition, which allow us to reveal some details that are not obvious for the Lagrangian formalism and which is required for a strict statistical description of a system of particles. General relativistic canonical equations of motion of scalar charged particles were formulated in the works of one of the Authors \cite{Ignatev1}, \cite{Ignatev2} (see also \cite{Ignat15}). Here we briefly list the main points of this formalism.  Normalized invariant volume element of the 8-dimensional phase space of a relativistic particle $\Gamma$, which is a vector bundle $\Gamma=P(X)\times X$ with a Riemannian base $X(g)$ and a vector layer $P(X)$ with respect to a pair of canonically conjugate dynamical variables $x^{i} $ (configuration coordinates) and $P_{i}$  (generalized momentum coordinates) is \cite{Bogolyub}:
\begin{eqnarray}\label{dG}
d\Gamma=\frac{\varrho}{(2\pi)^3}dXdP\equiv \frac{\varrho}{(2\pi)^3}dx^1dx^2dx^3dx^4dP_1dP_2dP_3dP_4,
\end{eqnarray}
where
\begin{eqnarray}
dX=\sqrt{-g}dx^1dx^2dx^3dx^4;\qquad dP=\frac{1}{\sqrt{-g}}dP_1dP_2dP_3dP_4
\end{eqnarray}
are the invariant elements of the volumes of the configuration and momentum spaces, respectively, and $\varrho$  is the degeneration factor (for particles with the spin $S$ $\varrho=2S+1$). Further, to shorten the notation, we will also use the phase coordinates of the same name $\eta_a$, $a=\overline{1,8}$:
\begin{equation}
\eta_i\equiv x^i,\quad \eta_{i+4}=P_i,\quad (i=\overline{1,4}),
\end{equation}
in which the expression for the element of the volume of the phase space (\ref{dG}) takes the simplest form:
\begin{equation}\label{dG1}
d\Gamma=\frac{\varrho}{(2\pi)^3}\prod\limits_{a=1}^8 d\eta_a.
\end{equation}
The canonical equations of motion of a relativistic particle in the phase space $\Gamma$ have the form (see e.g., \cite{Ignatev2}):
\begin{equation} \label{Eq1}
\frac{dx^{i} }{ds} =\frac{\partial H}{\partial P_{i} } ;\quad \quad \frac{dP_{i} }{ds} =-\frac{\partial H}{\partial x^{i} } ,
\end{equation}
where $H(x,P)$ - is a relativistically invariant Hamilton function, $u^i=dx^i/ds$ -- is a particle velocity vector.

Due to the antisymmetry of the canonical equations of motion (\ref{Eq1}) and the symmetry of the phase volume (\ref{dG}) with respect to the canonical variables $\{x^i,P_i\}$ the differential relation \cite{Bogolyub}, which is known in classical dynamics as {\it Liouville's theorem}, fulfills
\begin{equation} \label{Liuvill}
\frac{d\Gamma}{ds}=0,
\end{equation}.
According to this relaton, the phase volume of the world tube of particles is constant.

Calculating the total derivative of the function of dynamic variables $\Psi (x^{i} ,P_{k} )$, taking into account (\ref{Eq1}) we find:
\begin{equation} \label{Eq2}
\frac{d\Psi }{ds} =[H,\Psi ],
\end{equation}
where the invariant Poisson brackets are introduced:
\begin{equation} \label{Eq3}
[H,\Psi ]=\frac{\partial H}{\partial P_{i}} \frac{\partial \Psi }{\partial x^{i} } -\frac{\partial H}{\partial x^{i} } \frac{\partial \Psi }{\partial P_{i} } ;\qquad [H,\Psi]=-[\Psi,H].
\end{equation}
Let us notice that the Poisson bracket \eqref{Eq3} can be rewritten in an explicitly covariant form using an%
 {\it operator of covariant differentiation according to Cartan}, $\widetilde{\nabla}_i$, (covariant derivative in the bundle $\Gamma$ \cite{Cartan}, see, for example, \cite{Bogolyub})\footnote{ For the first time, covariant derivatives according to Cartan were introduced into relativistic statistics by A.A. Vlasov \cite{Vlasov}.}:
\begin{equation}\label{Cartan}%
\widetilde{\nabla}_i = \nabla_i +
\Gamma_{ij}^k P_k\frac{\partial}{\partial P_j},
\end{equation}
where $\nabla_i$ -- is an operator of covariant Ricci differentiation %
and $\Gamma^k_{ij}$ -- are Christoffel symbols of the second kind with respect to the metric $g_{ij}$ of the base  $X$. %
The operator $\widetilde{\nabla}$ is defined in such a way, that
\begin{equation}\label{9.11}%
\widetilde{\nabla}_iP_k \equiv 0
\end{equation}
and the following {\it symbolic} rule of differentiation of the functions \cite{Bogolyub} fulfills:
\begin{equation}\label{9.13}%
\widetilde{\nabla}_i\Psi(x,P) = \nabla_i[\Psi(x],P),
\end{equation}.
This rule means that in order to calculate the Cartan derivative of the function $\Psi(x,P)$ it suffices to calculate from it the usual covariant
derivative as if the momentum vector were covariantly constant. Due to this equality, the introduced operator is quite convenient %
for executing differential and integral operations in the phase space $\Gamma$. Thus, we write down the Poisson bracket \eqref{Eq3} in an explicitly covariant form:
\begin{equation}\label{H_Cart}
[H,\Psi ]\equiv \frac{\partial H}{\partial P_{i}} \widetilde{\nabla}_i\Psi-\frac{\partial \Psi}{\partial P_{i}} \widetilde{\nabla}_i H,
\end{equation}
Further, due to (\ref{Eq3}) the Hamilton function is the integral of the particle motion:
\begin{equation} \label{Eq4}
\frac{dH}{ds} =[H,H]=0,\Rightarrow H= \mathrm{Const}.
\end{equation}
The relation   (\ref{Eq4}) can be called the normalization relation. Due to the linearity of the Poisson bracket
any continuously differentiable function $f(H)$ is also a %
Hamiltonian function. The only way to introduce an invariant Hamilton function, %
which would be quadratic in the generalized momentum of a particle, in the presence of only gravitational and scalar fields, is the following::
\begin{equation} \label{Eq7}
H(x,P)=\frac{1}{2} \left[\psi(x)(P,P)-\varphi(x) \right],
\end{equation}
where $(a,b)$ here and further is the scalar product of the vectors $a$ and $b$ with respect to the base metric:
\begin{equation}(a,b)=g_{ik} a^{i} b^{k} ,\nonumber\end{equation}
and $\psi(x)$ and $\varphi(x)$ -- are certain scalar functions of scalar potentials.
Let us choose the zero normalization of the Hamilton function in the relation  \eqref{Eq4}:
\begin{equation}\label{Eq7a}
H(x,P)=\frac{1}{2} \left[\psi(x)(P,P)-\varphi(x) \right]=0,
\end{equation}
wherefrom we find:
\begin{equation}\label{Eq7b}
(P,P)=\frac{\varphi}{\psi},
\end{equation}
and from the first group of canonical equations of motion  (\ref{Eq1}) %
we obtain a relation between the generalized momentum and the particle velocity vector:
\begin{equation} \label{Eq10a}
u^{i} \equiv \frac{dx^{i} }{ds} =\psi P^{i} \Rightarrow P^{i} =\psi^{-1} u^{i} ,
\end{equation}
Substituting \eqref{Eq10a} into the normalization relation (\ref{Eq7b}),
we find:
\begin{equation}(u,u)=\psi\varphi.\nonumber\end{equation}
Therefore, to fulfill the normalization relation for the particle velocity vector
\begin{equation} \label{Eq11}
(u,u)=1
\end{equation}
it should be:
\[\psi\varphi=1 \Rightarrow \psi=\varphi^{-1},\]
-- thus, the invariant Hamilton function of a particle can be determined by just a single scalar function $\varphi(x)$. Taking into account the last relation, we write
Hamilton's function in final form:
\begin{equation}\label{Eq7 }
H(x,P)=\frac{1}{2} \left[\varphi^{-1}(x)(P,P)-\varphi(x) \right]=0,
\end{equation}
and from the canonical equations (\ref{Eq1}) we obtain the connection between the generalized momentum and the particle velocity vector:
\begin{equation}\label{Eq10}
P^i=\varphi \frac{dx^i}{ds}\equiv \varphi u^i.
\end{equation}
From the relation  (\ref{Eq7b}) it follows that the vector of the generalized impulse, as well as the vector of the velocity, is timelike:
\begin{equation}\label{Eq8}
(P,P)=\varphi^2\geqslant 0.
\end{equation}
Let us note a relation, which will be important in future and is a consequence of (\ref{H_Cart}), (\ref{Eq7}) и  (\ref{Eq8}):
\begin{equation} \label{Eq9}
[H,P^{k} ]=\nabla ^{k} \varphi \equiv g^{ik} \partial _{i} \varphi;
\end{equation}
where $\nabla^i\equiv g^{ik}\nabla_k$ -- is a symbol of a covariant derivative.

Thus, the following statement is true.

\begin{stat}\label{note_can1}
Within the framework of the canonical formalism, the motion of a scalar charged particle in a gravitational field is described by the phase trajectory  $x^i(s)$ in an 8-dimensional phase space, which is a vector bundle with a base $X$ and a layer $P(X)$ with an invariant element of volume $d\Gamma=dXdP$ \eqref{dG} which is conserved along the phase trajectories \eqref{Eq1}. The pair of phase coordinates of the particle $\{x^i,P_i\}$, configuration coordinates and generalized momenta, are canonically conjugated, which manifests itself, firstly, in the conjugate character of the equations of motion \eqref{Eq1}, and secondly, in the conservation of the phase volume \eqref{Liuvill} and, third, in the conservation of the invariant Hamilton function along the phase trajectories, -- \eqref{Eq4}. In this case, the invariant Hamilton function of a scalar charged particle is determined by just a single scalar function $\varphi(x)$ \eqref{Eq7 }, which depends on scalar potentials, and is equal to zero along the phase trajectories \eqref{Eq7a}. The relationship between the generalized momentum, $P_i$, and the particle velocity vector, $u^i=dx^i/ds$, is established by the relation \eqref{Eq10}.
\end{stat}

From the second group of canonical equations (\ref{Eq1}) пwe obtain the equations of motion in the Lagrangian formulation, which coincide with the above cited \eqref{du/ds}
\begin{equation} \label{Eq12}
\frac{d^{2} x^{i} }{ds^{2} } +\Gamma _{jk}^{i} \frac{dx^{j} }{ds} \frac{dx^{k} }{ds} =\partial _{,k} \ln |\varphi|\ \pi^{ik} ,
\end{equation}
 where $\pi^{ik}$ --  is the tensor of orthogonal projection onto the direction $u$ \eqref{pi_ik}. From the properties of the tensor \eqref{pi_iku^k} and the Euler equations (\ref{Eq12}) follows a strict consequence of orthogonality of the vectors of velocity and acceleration
\begin{equation} \label{Eq15}
g_{ik} u^{i} \frac{du^{k} }{ds} \equiv 0.
\end{equation}
Let us note that the Lagrangian equations of motion (\ref{Eq12}) are invariant with respect to the sign of the scalar function $\varphi(x)$:
\begin{equation} \label{Eq16a}
\varphi(x)\rightarrow -\varphi(x).
\end{equation}
The Hamilton function (\ref{Eq7 }) is also invariant under the transformation (\ref{Eq16a}) \emph{at its zero normalization}. Therefore, from the relations (\ref{Eq8}), (\ref{Eq10}), as well as the Euler equations (\ref{Eq12}) it follows that the square of the scalar $\varphi $ has the meaning of the square of the \textit{ dynamic mass of the particle, $m_{*} $, in scalar field}:

\begin{equation} \label{Eq16}
\varphi^2 =m_{*}^2 .
\end{equation}
The action function \eqref{S_m1} corresponds to the specified choice of the Hamilton function. This function formally coincides with the Lagrange function of a relativistic particle with a rest mass $m_*$ in a gravitational field (see, for example, \cite{Land}).

Let us return to the question of choice of the function of the dynamic mass of the particle. Now let the system have $n$ different scalar fields, $\Phi_r$, and each particle has $n$ corresponding fundamental scalar charges, $q_r$ ($r=\overline{1,n}$), which may include zero charges. The question about the choice of the function $m_{*} (\Phi_r)$ raises. Without specifying this function yet, let us note the following important circumstance. For the canonical equations of motion \eqref{Eq1} to admit the linear integral of motion \eqref{Eq1} according to \eqref{Eq2} it is necessary and sufficient that $[H,\Psi]=0$ which, in turn, is possible if and only if it is \cite{Ignat14_1}:
\begin{equation}\label{Lin_Int}
(\xi,P)=\mathrm{Const} \Leftarrow\!\!\Rightarrow \ \Lee{\xi}\varphi g_{ik}=0,
\end{equation}
where $\Lee{\xi}$ -- is a Lie derivative in the direction $\xi$ (see, e.g., \cite{Petrov}).

 Let us note that the linear integrals of motion \eqref{Lin_Int} have a transparent physical meaning: if $\stackunder{\alpha}{\xi}^i(x)$ is a spacelike vector $(\stackunder{\alpha}{\xi},\stackunder{\alpha}{\xi})<0$, then $\mathrm{C}_\alpha=(\stackunder{\alpha}{\xi},P)$ is a projection of the vector of the total momentum of the particle which is conserved along the direction $\stackunder{\alpha}{\xi}^i(x)$, while if $\stackunder{4}{\xi}^i(x)$ is a timelike vector  $(\stackunder{\alpha}{\xi},\stackunder{\alpha}{\xi})>0$, then $\mathrm{C}_4=(\stackunder{4}{\xi},P)$ is the conserved total energy of the particle.

Let us consider static fields $g_{ik} $ and $\Phi_r$, admitting a timelike Killing vector $\xi ^{i} =\delta _{4}^{i} $ and $\Phi(x^1,x^2,x^3)$, when the total energy of the charged particle is conserved,  $P_{4} =E_{0}=\mathrm{Const} >0$. Let's choose a frame of reference in which $g_{\alpha 4} =0, \alpha,\beta=\overline{1,3}$, so that the coordinate $x^{4}$ coincides with the world time $t$. Then from the relationship between the vector of the kinematic velocity $u^{i} $ and the vector of the total momentum of the particle $P_{i} $ \eqref{Eq10} it follows:
\begin{equation}\label{dt/ds}
P^i=\varphi u^i\Rightarrow u^4\equiv \frac{dt}{ds}=\frac{E_0}{\varphi} .
\end{equation}
Therefore, if we {\it like} to keep the same orientation of the world time $t$, and proper time $s$ (i.e., $u^4=dt/ds>0$), it is necessary to choose a mass function that would always remain non-negative: $m_{*} =|\varphi| >0$. However, such a conservative and, at first glance, correct approach used in the above cited works  \cite{Ignatev1}, \cite{Ignatev2}, \cite{Ignatev3}, \cite{Ignatev4}, contradicts the more fundamental principle of additivity of the Lagrange function. As is shown in \cite{Ignat14_1}, the negativity of the particle dynamic mass function does not lead to any contradictions at the level of microscopic dynamics, since \emph{the observed kinematic momentum of the particle}, which in our case exactly coincides with the total momentum, --
\begin{equation}\label{Pp}
p^i=m_*\frac{dx^i}{ds}\equiv P^i,
\end{equation}
just like the \emph{observed 3-dimensional velocity} $v^\alpha=dx^\alpha/dt\equiv  u^\alpha/u^4$, in contrast to unobservable kinematic 4-velocity of a particle, $u^i$,  conserves its orientation.  the choice of a linear mass function in \eqref{Eq7 } which coincides with \eqref{m_*(pm)}, corresponds to the principle of additivity of the action function. This choice also meets aesthetic criteria, since in this case the Hamilton function \eqref{Eq7 } нdoes not depend on the rest mass.

When choosing the dynamic mass in the form \eqref{m_*(pm)} the Hamilton function \eqref{Eq7 } and the normalization relation (\ref{Eq8}) for the generalized momentum take the form:
\begin{equation}\label{H,m}
H(x,P)=\frac{1}{2} \left[m_*^{-1}(x)(P,P)-m_*\right]=0,
\end{equation}
\begin{equation}\label{P_norm}
(P,P)=m^2_*.
\end{equation}
Let us note the identity laws, which would be useful further and are valid for the Hamilton function \eqref{H,m}:
\begin{equation}\label{nabla_H}
\widetilde{\nabla}_iH=-\nabla_i m_*,
\end{equation}
\begin{equation}\label{HPsi}
[H,\Psi]=\frac{1}{m_*}P^i\widetilde{\nabla}_i\Psi+\partial_i m_*\frac{\partial \Psi}{\partial P_i},
\end{equation}
where $\Psi(x,P)$ is an arbitrary function.

Thus, the following statement is fair.
\begin{stat}
From the point of view of the canonical formalism, the choice of a dynamic mass in the form \eqref{m_*(pm)} corresponds to the additivity principle of the Lagrange function of a scalar charged particle. In the case $m_*<0$ the relative orientation of the world and proper time of the particles $u^4=dt/ds<0$ changes , however, the orientation of the observed kinematic momentum of the particle \eqref{Pp}, as well as the orientation of the observed 3-dimensional velocity vector of the particle, $\mathbf{v}$, is conserved.
\end{stat}

\section{Lagrangian Formalism for a System of Scalar Charged Particles}
\subsection{Invariant and Tensor Functions of Singular Sources}

\qquad Since the questions related to the dynamic mass are rather delicate, below we provide a microscopic argumentation of the mathematical model of a plasma with scalar charged particles that is considered in this article. When formulating microscopic equations, it is necessary to correctly construct \emph{microscopic densities of singular field sources}.

Let us call a function $D(x_1|x_2)$ possessing following properties, an \emph{invariant with respect to non-degenerate transformations} ${x'}^i=\phi^i(x^1,\ldots,x^n)$ \emph{two-point }$\delta$ - \emph{Dirac function}, defined on $n$ - dimensional Riemann manifold:
\begin{equation}\label{D}
\int\limits_{X_2}D(x_1|x_2)F(x_2)dX_2=
\left\{\begin{array}{ll}
F(x_1), & x_1\in X_2;\\
0, & x_1 \not\in X_2,
\end{array}\right.
\end{equation}
($X_2\subset R_n$; $F(x)$ -- an arbitrary tensor field on $R_n$, $dX=\sqrt{-g} dx^1\ldots dx^n$ -- an invariant volume element $R_n$,
\begin{eqnarray}\label{sym_D}
D(x_1|x_2)=D(x_2|x_1);\\
\label{trans_D}
D(x'_1|x'_2)=D(x_1|x_2).
\end{eqnarray}
Along with the invariant Dirac $\delta$ - function, defined by the properties \eqref{D} -- \eqref{trans_D}, we can also consider the \emph{scalar density} $\Delta(x_1|x_2)$, which is usually called a Dirac $\delta$ - function (see \cite{Ignat_Delta}):
\begin{equation}\label{D->delta}
\Delta(x_1|x_2)=\frac{1}{\sqrt{-g}} D(x_1|x_2);\quad \Delta(x'_1|x'_2)=|J^{-1}(x_2)|\Delta(x_1|x_2),
\end{equation}
where $J$ -- is a Jacobian of transformation.

It can be shown (see \cite{Ignat_Delta}), that the invarian Dirac $\delta$ - function complies with the following symbolic rule of covariant differentiation:
\begin{equation}\label{dD}
\frac{\partial}{\partial x^i_1}D(x_1|x_2)=D(x_1|x_2)\frac{\partial}{\partial x^i_2}\Rightarrow
\stackrel{1}{\nabla}_i D(x_1|x_2) = D(x_1|x_2) \stackrel{2}{\nabla}_i,
\end{equation}
where $\stackrel{a}{\nabla}_i$ - is an operator of covariant differentiation in the point $x_a.$

The geometric image of a classical particle is a time-like world line $\Gamma_a:\ x^i=x^i(s_a)\equiv x^i_a$ along which a certain geometric object is given which characterizes its physical properties. Let us call this object a \emph{source}. A classical point particle in scalar fields is associated with only one tensor object - the velocity vector $u^i_a=dx^i_a/ds_a$, as well as scalars - scalar charges  $q^r_{(a)}$ and scalar dynamic mass $m_{(a)}$. Thus, the source of a classical particle can only have a structure of the form $\omega_a(s_a)=\{q^r_{(a)},m_{(a)}\}\times u^{i_1}\ldots u^{i_m}$. Let us define the density field of the source \cite{Ignat_Delta}
\begin{equation}\label{Omega_a}
\Omega_a(x)=\int\limits_{\Gamma_a} \omega_a D(x|x_a(s_a))ds_a,
\end{equation}
where the integration is carried out over the entire trajectory of the particle. Due to the definition of the invariant $\delta$- function,
the integration in \eqref{Omega_a} transfers the tensor properties of the object $\omega_a$ from the trajectory of the particle to the entire manifold $R_n$, specifying tensor field $\Omega_a(x)$.

If in some coordinate system $R_4$ can be represented as a direct product of a three-dimensional non-isotropic hypersurface $V_k$ and the congruence of the coordinate lines $x_k$, normal to it in each point, then in this coordinate system the invariant Dirac $\delta$- function can be
represented as the product of the invariant on the hypersurface $V_k$ 3-dimensional $\delta$ - function $D(\tilde{x}_1|\tilde{x}_2)$ and one-dimensional  $\delta$ - function $\delta(\stackunder{k}{x}\!\! _1|\stackunder{k}{x}\!\! _2)$
\begin{equation}\label{3.10}
D(x_1\,|\,x_2) = D( \tilde{x}_1|\tilde{x}_2)\delta(\stackunder{k}{x}\!\! _1|\stackunder{k}{x}\!\! _2)
\end{equation}
where $\tilde{x}$ - are coordinates on $V_k$. In this coordinate system, the volume element $R_4$ is also represented as the product $dX = dV_kdx_k$, where
$dV_k = \sqrt{-g(x)}d^{n-1}\tilde{x}$ - is the area element of the hypersurface $V_k$. In the future, we will often perform such an operation in a synchronous reporting system when the normal vector $k_i$ is timelike. The metrics $R_4$ has the following form in this frame of reference
\[
ds^2 = d\tau^2 + g_{\alpha \beta}dx^{\alpha}dx^{\beta}, \quad (\alpha ,\beta =\overline{1,3}).
\]
In this case, the $3$-dimensional hypersurface is spacelike, and its area element will be denoted by $dV$.

Let us consider the following source densities that have a simple physical meaning:
\begin{equation}\label{3.15}
n^i(x) = \sum n^i_a(x) = \sum\limits_a\int\limits_{\Gamma_a} u^i_a(s_a)D(x|x_a)ds_a
\end{equation}
-- is a particle number density vector (numeric vector),
\begin{equation}\label{3.16}
j^i_r(x) =\sum q^r_{(a)}n^i_a(x) = \sum\limits_a q^r_{(a)}\int\limits_{\Gamma_a} u^i_a(s_a)D(x|x_a)ds_a
\end{equation}
-- is a vector of current density of scalar charges.

Let's calculate the covariant divergences from these values. First, let us consider an expression of the form
\[\nabla_in^i_a(x) = \int\limits_{\Gamma_a}u^i_a(s_a)\frac{\partial D(x|x_a)}{\partial x^i}ds_a.\]
We represent the integral in this expression as a curvilinear integral of the 2nd type, taken along the entire trajectory
of a particle
\[\nabla_in^i_a(x) = \int\limits_{\Gamma_a}\frac{\partial {\cal D}(x\,|\,x_a)}{\partial x^i}dx^i_a\ ,\]
and take into account the symbolic rule \eqref{dD}, according to which the operator of differentiation should impact in our case on one. Thus, the following ratio always takes place
\begin{equation}\label{3.18}
\nabla_i n^i_a (x) = \int\limits_{\Gamma_a} u^i_a (s_a)
\frac{\partial {\cal D}(x\,|\, x_a)}{\partial x^i}ds_a = 0.
\end{equation}
This relation has the form of a conservation law and establishes
the obvious fact of the existence of a particle on its own
trajectory. As a consequence of (\ref{3.18}), the microscopic conservation laws are fulfilled:
\begin{equation}\label{3.19}
\nabla_in^i(x) = 0;
\end{equation}
\begin{equation}\label{3.20}
\nabla_ij_r^i(x) = 0.
\end{equation}

\subsection{Microscopic Field Equations with Singular Sources}\label{R4}
Taking into account the definition of the invariant $\delta$ - function \eqref{D} the action for the particle ``$a$''  \eqref{S_m1} can be rewritten as an integral over the volume of the Riemannian space $R_4\equiv X$:
\[S =-\int\limits_X dX \int\limits_{\Gamma_a} D(x|x_a) m_{(a)}(s_a)ds_a. \]
This allows us to write down the total action integral for the system ``scalar charged particles ($p$)'' +``scalar fields ($s$)'' +``gravitational field ($g$)'', in the form of an integral over $X$:
\begin{eqnarray}\label{S3}
S=S_p+S_s+S_g=-\int\limits_X dX\sum\limits_a \int\limits_{\Gamma_a}D(x|x_a)\ m_{(a)}(s_a)ds_a - \int\limits_X L_s dX -\frac{1}{16\pi}\int\limits_X(R+2\Lambda)dX,
\end{eqnarray}
where $\Lambda$ -- is the cosmological constant, $L_s$ -- is the Lagrange function of noninteracting scalar fields
\begin{eqnarray} \label{Ls}
L_s=\sum\limits_r L_{(r)}= \frac{1}{16\pi}\sum\limits_r(e_r g^{ik} \Phi_{r,i} \Phi _{r,k} -2V_r(\Phi_r)),
\end{eqnarray}
\begin{eqnarray}
\label{Higgs}
V_r(\Phi_r)=-\frac{\alpha_r}{4} \left(\Phi_r^{2} -\frac{m_r^{2} }{\alpha_m}\right)^{2}
\end{eqnarray}
-- is a potential energy of corresponding scalar fields, $\alpha$ and $\beta$ -- are constant of their self-action, $m_r$  -- their quanta masses, $e_r=\pm 1$ -- are indicators (sign ``+'' corresponds to classical scalar fields, sign ``-'' corresponds to phantom ones).

Вычислим вариацию действия \eqref{S3} по динамическим переменным частиц, $\delta_a S=\delta_aS_p$. При этом надо учитывать свойство дифференцирования инвариантной $\delta$ - функции \eqref{dD}, вследствие которого операция вариации также переносится на функции справа $\delta D(x|y)F(y)=D(x|y)\delta F(y)$. В результате получим:
\[ \delta_aS=-\int\limits_X dX\sum\limits_a \int\limits_{\Gamma_a}D(x|x_a)\delta_a(m_{(a)}(s_a)ds_a).\]
Equating the variation $\delta_pS $ to zero, we obtain the equations of motion of scalar charged particles in the form \eqref{du/ds_m}
\begin{equation}\label{du_a/ds_m}
\frac{\delta u_a^i}{\delta s_a}=\partial_k\ln|m_{(a)}|\ \pi^{ik}(u_a),
\end{equation}

Calculating now the variation of the action of particles in scalar fields $\delta_r S_p$ taking into account the definition of the dynamic mass of particles \eqref{m_*(pm)} in a similar way, we obtain:
\[\delta_r S_p=-\int\limits_X dX\sum\limits_a q^r_{(a)}\int\limits_{\Gamma_a}D(x|x_a)q^r_{(a)}\delta\Phi_r  ds_a. \]

Varying the actions of scalar fields with respect to scalar fields, we find the total variations $\delta_rS=\delta_r S_p+\delta_r S_s$
\[\delta_r S= -\int\limits_X dX \delta\Phi_r \left(\sum\limits_a q^r_{(a)} \int\limits_{\Gamma_a}D(x|x_a) ds_a+\frac{1}{8\pi}\bigl(e_r\Box\Phi_r+V'_{\Phi_r}\bigr) \right). \]
Thus, we obtain microscopic equations of scalar fields:
\begin{equation}\label{Box(Phi)}
e_r\Box \Phi + V'_{\Phi_r} = -8\pi \sum\limits_a q^r_{(a)}\int D(x|x_a)ds_a;\quad (r=\overline{1,N}),
\end{equation}
where $\Box$ -- is the D'Alembert operator on the $g_{ik}$.

Let us write down the equations of scalar fields \eqref{Box(Phi)} in a more compact form:
\begin{equation}\label{Box(Phi)=sigma}
e_r\Box \Phi + V'_{\Phi_r} = -8\pi\sigma^r,\quad (r=\overline{1,N}),
\end{equation}
where the \emph{microscopic densities of the scalar charge of a system of particles with respect to the scalar field} $\Phi_r$ are introduced:
\begin{equation}\label{sigma_r}
\sigma^r\equiv \sum\limits_a \sigma^r_{(a)}= \sum\limits_a q^r_{(a)}\int D(x|x_a)ds_a,
\end{equation}
possessing the following symmetry property with respect to scalar charges
\begin{equation}\label{sigma(-q)}
\sigma^r_{(a)}(-q^r_{(a)})=-\sigma^r_{(a)}(q^r_{(a)}),\quad (r=\overline{1,N}).
\end{equation}

Finally, we calculate the variation of the action \eqref{S3} with respect to the gravitational field. When calculating the variation of the action of particles, we obtain:
\[ \delta_gS_p=-\int\limits_X dX\sum\limits_a \int\limits_{\Gamma_a}D(x|x_a)m_{(a)}(s_a)\delta_g ds_a.\]
Thus, we find, taking into account the ratio of the normalization of the velocity vector $u_a,u_a)=1$:
\[\delta_g ds_a=\frac{\partial}{\partial g_{ik}}\sqrt{g_{lm} u^l_a u^m_a}ds_a\delta g_{ik}=\frac{1}{2}u^i_a u^k_a \delta g_{ik} ds_a.\]
The result of calculating this variation on the action functions of the scalar fields and the gravitational field is known, and as a result we obtain the microscopic Einstein equations:
\begin{equation}\label{EqEinst_micro}
G^i_k\equiv R^i_k-\frac{1}{2}R\delta^i_k=8\pi(T^i_{(p) k}+T^i_{(s) k})+\Lambda\delta^i_k,
\end{equation}
where the \emph{microscopic energy - momentum tensor of scalar charged particles}, $T^i_{(p) k}$, is introduced --
\begin{equation}\label{T_p_m}
T^{ik}_{(p)}(x) = \sum\limits_a T^{ik}_a(x) = \sum\limits_a \int\limits_{\Gamma_a} m_{(a)}(s_a) u^i_a(s_a)u^k_a(s_a)D(x|x_a)ds_a,
\end{equation}
and energy - momentum tensor of scalar fields,  $T^i_{(s)k}$, --
\begin{eqnarray}\label{T_s}
T^i_{(s)k}=\frac{1}{16\pi }\sum\limits_r\bigl(2\Phi^{,i}_{r} \Phi _{r,k} -\delta^i_k\Phi _{r,j} \Phi _r^{,j} +2V_r(\Phi_r)\delta^i_k \bigr).
\end{eqnarray}

\begin{note}\label{note_sign} Let us note a very important direct consequence of the fundamental principle of additivity of the action function: Einstein's equations are determined by the tensor of energy - momentum of particles with dynamic masses of particles $m_{(a)}$ \eqref{m_*(pm)}. This consequence is not quite usual, since, at first glance, it can lead to negative values of the energy density of matter at $m_{(a)}<0$. In this regard, taking into account  \eqref{Pp} let us transform the expression for the tensor of energy - momentum of scalar charged particles \eqref{T_p_m}
\begin{eqnarray}\label{T_p_m2}
T^{ik}_{(p)}(x) = \sum\limits_a \int\limits_{\Gamma_a} P^i_a(s_a)u^k_a(s_a)D(x|x_a)ds_a\equiv \sum\limits_a \int\limits_{\Gamma_a} P^i_a(s_a)D(x|x_a)dx^k_a,
\end{eqnarray}
We see that the radical, at first glance, decision on the choice of the sign of the dynamic mass \eqref{m_*(pm)} leads to the correct sign for the energy - momentum tensor of scalar charged particles, because of \eqref{Pp}. Note that if we discard the additivity principle of the Lagrange function as erroneous in our case and replace in the Lagrange function of a scalarly charged particle \eqref{S_m1} its dynamic mass by the total one   $m_*\to |m_*|$,then the equations of motion of the particle \eqref{du/ds} will not change. In this case, however, the change of energy - momentum tensor of scalar charged particles \eqref{T_p_m} and density of scalar charges \eqref{sigma_r}, -- it is required to perform a substitution $m_*\to |m_*|$ in the energy - momentum tensor and , а in the scalar charge density, introduce the sign function $\mathrm{sign}(m_{(a)})$ as a multiplier under the integral sign. As a result, the sign function $\mathrm{sign}(m_{(a)})$ has to be introduced as a multiplier in the integrand \eqref{T_p_m2}, which exactly leads to negative values for particles energy density.
\end{note}

Repeating similar calculations for the divergence of the energy-momentum tensor of particles \eqref{T_p_m}, taking into account the equations of motion \eqref{du_a/ds_m}, we obtain:
\begin{eqnarray}\label{d_kT^ik_p}
\nabla_kT^{ik}_{(a)}(x) = \int\limits_{\Gamma_a} D(x|x_a)u^k_a {\stackrel{a}{\nabla}} _k m_{(a)}(s_a)u^i_a(s_a) ds_a\Rightarrow
\nabla_kT^{ik}_p=\sum\limits_a \int\limits_{\Gamma_a} D(x|x_a){\stackrel{a}{\nabla}}\ \!^i m_{(a)}(s_a) ds_a
\end{eqnarray}
Taking into account the symmetry of the invariant $\delta$ - function \eqref{sym_D}, the definition of the dynamic mass of the particle \eqref{m_*(pm)} and \eqref{sigma_r}, let us rewrite \eqref{d_kT^ik_p} in a more convenient form:
\begin{equation}\label{div_sing_T_p}
\nabla_kT^{ik}_{(p)}=\sum\limits_a \sum\limits_r q^r_{(a)} \nabla^i\Phi_r \int\limits_{\Gamma_a} D(x|x_a) ds_a\Rightarrow
\nabla_kT^{ik}_{(p)}=\sum\limits_r \sigma^r \nabla^i \Phi_r.
\end{equation}

It is easy to verify that, due to \eqref{div_sing_T_p}, \eqref{Box(Phi)} and \eqref{T_s} the law of conservation of the total tensor of energy - momentum of the system ``scalar charged particles' + ``scalar fields'' is automatically fulfilled
\begin{equation}\label{nabla(Tp+Ts)}
\nabla_k T^{ik}=0 ; \qquad(T^{ik}=T^{ik}_{(p)}+T^{ik}_{(s)}),
\end{equation}
where energy - momentum tensor of scalar charged particles, $T^{ik}$, and energy - momentum tensor of scalar fields, $T^{ik}_{(s)}$, are described by formulas \eqref{T_p_m} and \eqref{T_s}, correspondingly.

\section{Invariant Distribution function, Macroscopic averages, and Transport Equations in the General Relativistic Kinetic Theory of Scalar Charged Plasma}
\qquad In this article, we will consider a macroscopic model of a scalar charged plasma based on the general relativistic kinetic and statistical theory. The foundations of the general relativistic kinetic and statistical theory were laid in 60's in the works by E.Tauberg - J.W. Weinberg \cite{taub}, N.A.Chernikov (see, e.g. \cite{chern}), A.A. Vlasov \cite{Vlasov} and others. Scalar fields were introduced into general relativistic statistics and kinetics in the early 80's in the works by one of the Authors  \cite{Ignatev1}, \cite{Ignatev2}, \cite{Ignatev3}, \cite{Ignatev4}. Further, in the works \cite{Ignat14_1,Ignat14_2,Ignat15} a mathematical model of the statistical system of scalar charged particles, based on the microscopic description and the subsequent procedure of transition to the kinetic and hydrodynamic models, was formulated. The already cited works \cite{Ignat15} and \cite{Ignat15_2} contained a correct generalization of the general relativistic kinetic theory to the sector of phantom fields and negative dynamic masses.

\subsection{Invariant Relativistic Distribution Function of Identical Particles}
The general formalism of invariant distribution functions was developed in the works \cite{Bogolyub,Yubook1}. To take into account the possibility of a negative sign of the dynamic mass of particles, it is necessary to carefully apply this formalism to the considered case. To determine macroscopic averages in a relativistic phase space, it is necessary to determine on its basis \emph{a unit timelike field of macroscopic observers}, $U_i(x) :(U,U)=1$, according to whose clock the synchronization of the acts of measurement of individual particles will be carried out. This timelike field, in turn, defines a spacelike three-dimensional surface, $V_3$, offsets along which, $\delta x^i$, are orthogonal to the given field:
\begin{equation}\label{Sigma}
V_3:\; \delta x^iU_i=0,
\end{equation}
and the offsets along this field $dx^i$ determine the synchronized proper time $\tau$ of the observers:
\begin{equation}\label{tau}
\frac{dx^i}{d\tau}=U^i \Leftrightarrow \frac{dx^i}{d\tau}U_i=1\Rightarrow d\tau=dx^iU_i.
\end{equation}
Thus, in the macroscopic reference frame of observers:
\begin{equation}\label{dXt}
X=V\times T \Rightarrow dX=dVd\tau.
\end{equation}

In this case, the connection of the proper time of the particle, $s$, with the synchronized proper time $\tau$ of observers at each point of the configuration space is established by the relation:
\begin{equation}\label{dtauds}
\frac{d\tau}{ds}=U_i\frac{dx^i}{ds}.
\end{equation}
Taking into account now the relation (\ref{Pp}), which is a consequence of the canonical equations (\ref{Eq1}), we finally obtain:
\begin{equation}\label{dsdtau}
\frac{d\tau}{ds}=m_*^{-1}(U,P)\Rightarrow \frac{1}{m_*(s)}\frac{ds}{d\tau}=\frac{1}{(U(\tau),P(s))}.
\end{equation}
The relation (\ref{dsdtau}) can be considered as a differential equation with respect to the function $s(\tau)$, solving which, we can determine the relationship between the proper time of the particle and the time, measured by the clocks of synchronized observers::
\begin{equation}\label{s(tau)}
s=s(\tau).
\end{equation}

Further, in the general relativistic kinetic theory, a system of particles is described by the distribution function $F_a(x,P)$, in canonically conjugate dynamical variables $\{x^i,P_i\}$,  which invariant in the 8-dimensional phase space $\Gamma= P(X)\times X$.
The invariant 8-dimensional distribution function of identical particles $F(x,P)$ is introduced in the following way \cite{Bogolyub}. Let the phase trajectory of a particle, determined by the canonical equations \eqref{H_Cart} in the phase space $\Gamma$ is:
\begin{equation}\label{x(s),P(s)}
x^i=x^i(s);\quad P_i=P_i(s)\Rightarrow \eta_a=\eta_a(s)\; (a=\overline{1,8}),
\end{equation}
where $s$ is a proper time of a particle.
Then the number of particles registered by observers in the region $d\Gamma$ of the phase space, %
can be defined as \cite{Bogolyub}:
\begin{equation}\label{dN}
dN(\tau)=F(x,P)\delta(s-s(\tau))d\Gamma.
\end{equation}
Note that the number of particles is a scalar, depending, however, on the choice of the field of observers $U^i$, i.e., on the choice of the frame of reference in the Riemannian space $X$,
while the 8-dimensional distribution function itself $F(x,P)$, introduced by the relation (\ref{dN}), is invariant in the phase space $\Gamma$. Note also that it is impossible to give another definition of the invariant distribution function in the 8-dimensional phase space. All previously introduced definitions of this function are special cases (\ref{dN}), implemented in the selected reference systems.

\subsection{Macroscopic Averages of the Dynamic Functions}
The definition of the invariant distribution function (\ref{dN}), and other dynamic functions together with it, is the first key point of the relativistic kinetic theory, which depends on the sign of the dynamic mass of particles. Therefore, the determination of the appropriate operations requires special care. Let $\psi(x,P)\equiv\psi(\eta)$ be some scalar function of dynamic variables. Then, according to (\ref{dN}) its macroscopic average $\Psi(\tau)$ in range $\Omega \subset\Gamma$ is determined as follows:
\begin{eqnarray}\label{Psi(tau)dG}
\Psi(\tau)=\int\limits_\Omega \psi(\eta(s))dN=\int\limits_\Omega F(\eta(s))\psi(\eta(s))\delta(s-s(\tau))d\Gamma.
\end{eqnarray}
Assuming further, in accordance with \eqref{Sigma}, \eqref{tau} $X=V\times T$ $\Rightarrow dX=dVdt$, let us write the expression (\ref{Psi(tau)dG}) in the explicit form:
\begin{eqnarray}\label{Psi(tau)dVt}
\Psi(\tau)=\frac{2S+1}{(2\pi)^3}\int\limits_V dV\int\limits_T dt \int\limits_{P(X)} dP \psi(\eta) F(\eta)\psi(\eta)\delta(s-s(\tau))
\end{eqnarray}
To integrate (\ref{Psi(tau)dVt}) over $t$ let us take into account the relationship between  $t(s)$ (\ref{Pp}) and %
 $\tau(s)$ (\ref{s(tau)}) and the properties of Dirac $\delta$-function:
\begin{eqnarray}\label{int_s}
\delta(s-s(\tau))dt=\left|\frac{d\tau}{ds}\right|\delta(t-\tau)ds\equiv |m_*|^{-1}(U,P)\delta(t-\tau)dt.
\end{eqnarray}
When deriving (\ref{int_s})we took into account the fact that the orientation of the generalized momentum, in contrast to the orientation of the kinematic velocity vector, does not depend on the sign of the effective mass. Taking now into account (\ref{int_s}) in (\ref{Psi(tau)dVt}) and integrating over the time coordinate, we finally get:
\begin{eqnarray}\label{Psi_7}
\Psi(\tau)=\frac{2S+1}{(2\pi)^3}\int\limits_V   \frac{U_i dV}{|m_*|} \int\limits_{P(X)} P^i dP \psi(\eta)F(\eta)\psi(\eta).
\end{eqnarray}
In particular, assuming $\psi=1$, we obtain the total number of particles in range $V$:
\begin{eqnarray}\label{N}
N(V)=\frac{2S+1}{(2\pi)^3}\int\limits_V  \frac{U_i dV}{|m_*|} \int\limits_{P(X)} P^i F(\eta)dP\equiv \int\limits_V  (U,n) dV,
\end{eqnarray}
where a {\it particle flux density vector} is introduced:
\begin{equation}\label{ni}
n^i(x)=\frac{2S+1}{(2\pi)^3|m_*|}\int\limits_{P(X)} P^i F(\eta)dP.
\end{equation}

Further, due to the generalized momentum normalization relation (\ref{P_norm}) the invariant 8-dimensional distribution function $F(x,P)$ is singular on a mass surface. Therefore, it is necessary to introduce a distribution function $f(x,P)$ that is nonsingular on this surface with a help of the relation:
\begin{equation}\label{f}
F(x,P)=\delta(H(x,P))f(x,P).
\end{equation}
Calculating the relation
\begin{equation}
F(x,P)dP\equiv\!\!\! \frac{1}{\sqrt{-g}}dP_1dP_2dP_3dP_4 \delta(H(x,P))f(x,P)\nonumber
\end{equation}
by means of Dirac $\delta$-function's properties and the explicit form of the Hamilton function \eqref{H,m}, we find:
\begin{eqnarray}\label{F,f}
F(x,P)dP=\frac{1}{\sqrt{-g}}dP_1dP_2dP_3  \frac{|m_*|}{P^4_+}\delta(P_4-P_4^+)f(x,P) \equiv |m_*|\delta(P_4-P_4^+)f(x,P)dP_0 ,
\end{eqnarray}
where $P_4^+\equiv P_4$ is a positive есть positive root of the normalization equation  (\ref{P_norm}),
а
\begin{equation}\label{dP_0}
dP_0=\frac{1}{\sqrt{-g}}\frac{dP_1dP_2dP_3}{P^4}
\end{equation}
is a volume element of three-dimensional momentum space.

Let us note that due to the antisymmetry of the Poisson bracket \eqref{Eq3} and \eqref{Eq4} the following relation holds for any function $G(H)$:
\begin{equation}\label{G(H)F}
\bigl[H,G(H)F(x,P)\bigr]= G(H)\bigl[H,F(x,P)\bigr],
\end{equation}
in particular, --
\begin{equation}
[H,\delta(H)f(x,P)]=\delta(H)[H,f(x,P)].
\end{equation}
As a result, the formulas for macroscopic averages \eqref{Psi_7} can be written in terms of {\it even-dimensional nonsingular distribution function} $f(x,P)$ in the following way:
\begin{eqnarray}\label{Psi_6}
\Psi(\tau)=\frac{2S+1}{(2\pi)^3}\int\limits_V  U_i dV\!\!\!
 \int\limits_{P_+(X)}\!\! P^i \psi(\eta)f(\eta)dP_0 ,
\end{eqnarray}
where it is necessary to substitute the positive root of the mass surface equation instead of $P_4$, and $P_+$ is an upper part of the pseudosphere of the mass surface. Thus, when passing to the 7-dimensional distribution function, the explicit dependence on the dynamic mass in these formulas disappears

As a result, the following symbolic rule which is being understood in a sense of of integration over the corresponding phase volumes, is valid:
\begin{eqnarray}\label{psi_rule}
\psi(\eta) F(\eta)\delta(s-s(\tau))d\Gamma\rightarrow \psi(\tilde{\eta})f(\tilde{\eta})(U,P)dVdP_0,
\end{eqnarray}
where $\tilde{\eta}$ are dynamic variables on the 6-dimensional subspace $\Gamma_0(\tau)=V\times P_0\subset \Gamma$
with the volume element:
\begin{equation}\label{G0}
d\Gamma_0=dVdP_0.
\end{equation}
In particular, for the {\it flux density vector of the number of particles}%
\footnote{Согласно J. Synge \cite{Sing}.} \eqref{ni} %
we obtain from \eqref{Psi_6}:
\begin{equation}\label{ni1}
n^i(x)=\frac{2S+1}{(2\pi)^3}\int\limits_{P_0(X)} P^i f(\eta)dP_0 .
\end{equation}

\subsection{The Macroscopic Conservation Laws }
As was noted above, in the cited works of one of the Authors \cite{Ignatev1,Ignatev2,Ignatev3,Ignatev4}, \cite{Ignat14_1,Ignat14_2,Ignat15}  the kinetic theory of scalar charged plasma was formulated. In particular, it was done in the work \cite{Ignat15} in generalization to the sector of phantom scalar fields and negative dynamic masses. The transport equations are strict macroscopic consequences of the kinetic theory\footnote{The transport equations are integro-differential consequences of kinetic equations, the description of which would take significant volume the article. For details, including the generalization to T-non-invariant interactions, please see in \cite{Ignat15}.}, including the conservation law of some vector current corresponding to the microscopic conservation law in reactions of a certain fundamental charge ${\rm G}$ (if there exists such a conservation law) -- %
\begin{equation}\label{1}
\nabla_i\sum\limits_a q^r_{(a)} n^i_{(a)}=0,
\end{equation}
as well as the conservation laws of the statistical system's energy - momentum:
\begin{equation}\label{2}
\nabla _{k} T_{p}^{ik} -\sum\limits_r\sigma^r\nabla ^{i} \Phi_r =0,
\mathrm{}\end{equation}
 where $n^i_a$ -- is a number vector, $T^{ik}_p$ -- is a tensor of particles' energy - momentum; $\sigma^r$ -- is a density of scalar charges with respect to the field $\Phi_r$ \cite{Ignat14_2}, so that
\begin{equation}\label{2a}
T^{ik}_p=\sum\limits_a T^{ik}_{(a)}; \quad \sigma^r=\sum\limits_a \sigma^r_{(a)};
\end{equation}
\begin{equation}\label{Tpl}
T^{ik}_{(a)}= \frac{2S+1}{(2\pi )^{3}}\int\limits_{P_0} f_a(x,P)P^iP^kdP_0;
\end{equation}
\begin{equation} \label{sigma_macr}
\sigma^r_{(a)} =\frac{2S+1}{(2\pi )^{3} } m_{(a)} q^{(r)}_{(a)}\int\limits_{P_0} f_a(x,P)dP_0.
\end{equation}
Further, calculating the divergence from the total energy-momentum tensor of the ``plasma+scalar fields'' system according to \eqref{T_s}, \eqref{2} and equating it to zero, we obtain
\[\sum\limits_r \nabla^i \Phi_r\biggl(e_r\Box\Phi_r+V'_{\Phi_r}+8\pi \sigma^r\biggr)=0,\]
whence, under the condition of functional independence of scalar fields, we obtain macroscopic equations of scalar fields with sources:
\begin{equation}\label{macr_eqs}
e_r\Box\Phi_r+V'_{\Phi_r}=-8\pi \sigma^r,\quad (r=\overline{1,N}).
\end{equation}

\begin{stat}\label{micro-macro}
Comparison of formulas for microscopic  \eqref{3.15}, \eqref{sigma_r}, \eqref{T_p_m} and corresponding macroscopic tensor densities \eqref{ni}, \eqref{Tpl}, \eqref{sigma_macr} with an account of the relation \eqref{Pp} reveals their full compliance accurate to nonsingular operation of averaging \eqref{Psi_6}. При In this case, microscopic and macroscopic conservation laws for the corresponding quantities  \eqref{3.18} $\leftrightarrow$ \eqref{1}, \eqref{div_sing_T_p} $\leftrightarrow$ \eqref{2}, as well as the equations of scalar fields with sources \eqref{Box(Phi)=sigma} $\leftrightarrow$ \eqref{macr_eqs} fully match. Thus, the introduced operations of calculating the macroscopic averages of dynamic functions are strictly justified on the microscopic dynamic level.

\end{stat}

\subsection{Local Thermodynamic Equilibrium}
 At sufficiently strong interparticle interaction, when the average free path of particles becomes much smaller than the characteristic size of the statistical system, or the free path is much less than the characteristic evolution time of the statistical system, the \emph{ocal thermodynamic equilibrium, LTE} is established in the statistical system. Under the local thermodynamic equilibrium (LTE) conditions, the statistical system is isotropic and is described by locally equilibrium distribution functions\footnote{Details see in \cite{Yubook1}}:
\begin{equation}\label{8_0}
f^0_a=\displaystyle \biggl[\exp\biggl(\frac{-\mu_{(a)}+(u,p)}{\theta}\biggr)\pm 1\biggr]^{-1},
\end{equation}
where $\mu_{(a)}$ -- is a chemical potential, $\theta$ -- is a local temperature, $u^i$ -- is a unit timelike  \emph{ector of the macroscopic kinematic velocity of the statistical system}, the sign ``$+$'' corresponds to fermions, ``$-$'' -- to bosons.  Macroscopic moments with respect to the isotropic distribution function \eqref{8_0} take the form of the corresponding moments of an ideal fluid for each of the components \cite{Ignat14_1}:
\begin{equation}\label{3}
n^i_a=n_a u^i,
\end{equation}
\begin{equation}\label{Tp_a_macr}
T^{ik}_a=(\varepsilon_{(a)}+p_{(a)}) u^iu^k-p_{(a)}g^{ik},
\end{equation}
where
\begin{equation}\label{5}
(u,u)=1.
\end{equation}
Since the vector $u^i$ is the eigenvector $T^{ik}_a$, corresponding to the eigenvalue $\varepsilon_{a}$, it is also the \emph{vector of the macroscopic dynamic velocity of the statistical system} \cite{Sing}.  The normalization relation  (\ref{5}) implies the well-known identity
\begin{equation}\label{6}
u^k_{~,i}u_k\equiv 0,
\end{equation}
which allows us to reduce the conservation laws of energy - momentum  (\ref{2})  to the form of the equations of ideal hydrodynamics
\begin{eqnarray}\label{2a}
(\varepsilon_p+p_p)u^i_{~,k}u^k=(g^{ik}-u^iu^k)\biggl(p_{p,k}+\sum\limits_r\sigma^r\Phi_{r,k}\biggr);\\
\label{2b}
\nabla_k[(\varepsilon_p+p_p)u^k]=u^k\biggl(p_{p,k}+\sum\limits_r\sigma^r\Phi_{r,k}\biggr),
\end{eqnarray}
and the fundamental charge conservation law $\mathrm{Q}$ \eqref{1} to the form
\begin{equation}\label{2c}
\nabla_k \rho^ru^k=0,
\end{equation}
where
\begin{equation}\label{n^r}
\rho^r\equiv \sum\limits_a q^r_{(a)} n_{(a)}
\end{equation}
-- \emph{кинематическая плотность скалярного заряда} статистической системы по отношению к скалярному полю $\Phi_r$.

Next, we calculate the macroscopic scalar densities $ n_{(a)}$, $\varepsilon_{(a)}$, $p_{(a)}$ и $\sigma^r_{a)}$ using the definitions \eqref{ni},
\eqref{Tpl}, \eqref{sigma_macr}, \eqref{3} and \eqref{Tp_a_macr}. Contracting the relations \eqref{3} with the velocity vector $u_i$, and the relations  \eqref{Tp_a_macr}, in sequence, with $u_iu_k$ and $\pi_{ik}(u)$ (see. \eqref{pi_ik}) with an account of the relation of the velocity vector's normalization \eqref{5}, we find the expressions for the macroscopic scalars:
\begin{eqnarray}\label{n,e,p}
 n_{(a)}=u_i n^i_{(a)}\quad  \varepsilon_{(a)}=u_iu_kT^{ik}_{(a)};\quad p_{a)}=\frac{1}{3}\pi_{ik}(u)T^{ik}_{(a)}.
\end{eqnarray}
Thus, proceeding to the integration over the three-dimensional space of momenta in the formulas \eqref{ni}, \eqref{Tpl}, \eqref{sigma_macr}, and taking into account the normalization relation for the generalized momentum \eqref{P_norm}, we find expressions for the macroscopic scalars:
\begin{eqnarray}
n_{(a)}=\frac{2S+1}{(2\pi)^3}\int\limits_{P_0} (u,P) f^0_{a}(x,P)\frac{dP_1dP_2dP_3}{\sqrt{-g}P^4};\nonumber\\
\varepsilon_{(a)}=\frac{2S+1}{(2\pi )^{3}}\int\limits_{P_0} (u,P)^2 f^0_a(x,P)\frac{dP_1dP_2dP_3}{\sqrt{-g}P^4};\nonumber\\
p_{(a)}=\frac{2S+1}{3(2\pi )^{3}}\int\limits_{P_0} \bigl[m^2_{(a)}-(u,P)^2\bigr] f^0_a(x,P)\frac{dP_1dP_2dP_3}{\sqrt{-g}P^4};\nonumber\\
\sigma^r_{(a)} =\frac{2S+1}{(2\pi )^{3} } m_{(a)} q^{r}_{(a)}\int\limits_{P_0} f_a(x,P)\frac{dP_1dP_2dP_3}{\sqrt{-g}P^4}.\nonumber
\end{eqnarray}
Перейдем в интегралах этих формул к сферической системе координат пространства импульсов в локальном репере $u^i=\delta^i_4$:
\begin{eqnarray}\label{sp_coord_P}
P_{(1)}=|m_{(a)}|z\cos\phi\cos\theta,\; P_{(2)}=|m_{(a)}|z\sin\phi\cos\theta,\; P_{(3)}=|m_{(a)}|z\sin\theta;\; P_{(4)}=|m_{(a)}|\sqrt{1+z^2};  \nonumber\\
 z\in [0,+\infty);\; \theta\in \biggl[-\frac{\pi}{2},\frac{\pi}{2}\biggr];\; \phi\in [0,2\pi];\; |J_p|=|m^3_{(a)}|z^2\cos\theta;\\ dP_0\equiv\frac{dP_1dP_2dP_3}{\sqrt{-g}P^4}=\frac{m^2_{(a)}}{\sqrt{1+z^2}}z^2\cos\theta dz d\theta d\phi,\nonumber
\end{eqnarray}
 where $P_{(k)}$ are orthoframe components of a generalized momentum, $J_p$ -- is a Jacobian of transformation to the spherical frame of reference. Let us notice, that $m_{(a)}$ is included in all formulas for microscopic scalars without the modulus sign which is the consequence of the correct definition of scalar densities. Thus, carrying out integration by angular momentum variables \eqref{sp_coord_P} in the formulas for macroscopic scalars, we finally get:
\begin{eqnarray}\label{6a}
n_{(a)}=|m_{(a)}|^3\frac{(2S+1)}{2\pi^2}\int\limits_{0}^\infty \frac{z^2dz}{\exp(-\gamma_{(a)}+\lambda_{(a)}\sqrt{1+z^2})\pm1};\\
\label{6b}
\varepsilon_{(a)}=m^4_{(a)}\frac{2S+1}{2\pi^2}\int\limits_{0}^\infty \frac{\sqrt{1+z^2}z^2dz}{\exp(-\gamma_{(a)}+\lambda_{(a)}\sqrt{1+z^2})\pm1};\\
\label{6c}
p_{(a)}=m_{(a)}^4\frac{2S+1}{6\pi^2}\int\limits_{0}^\infty \frac{1}{\sqrt{1+z^2}}\frac{z^4dz}{\exp(-\gamma_{(a)}+\lambda_{(a)}\sqrt{1+z^2})\pm1};\\
\label{6e}
\sigma^r_{(a)} =m^3_{(a)}q^{r}_{(a)}\frac{2S+1}{(2\pi)^{3}}
\int\limits_{0}^\infty\frac{1}{\sqrt{1+z^2}}\frac{z^2dz}{\exp(-\gamma_{(a)}+\lambda_{(a)}\sqrt{1+z^2})\pm1},
\end{eqnarray}
where $\varepsilon_p=\sum\varepsilon_a$, $p_p=\sum p_a$, $\sigma^r=\sum\sigma^r_{(a)}$,  $\lambda _{(a)} =|m_{(a)}| /\theta $, $\gamma_{(a)}=\mu_a/\theta$, $S$ -- is a spin of particles. Thus, the absolute magnitude of the dynamic mass is included in the final formulas for the equilibrium macroscopic scalars by only means of the equilibrium distribution function through the definition of energy $P_4$.

Let us note that according to definitions \eqref{n^r}, \eqref{6a} and \eqref{6e} the kinematic density of a scalar charge $\rho^r$ \eqref{n^r}, determining its conservation, does not coincide with the scalar charge's scalar density $\sigma^r$ which is a source of a scalar field $\Phi_r$. These densities are match by absolute magnitude only for non-relativistic plasma, where average values of the momentum are small: ${\overline{z}}^2\to0$. In addition, for $m_{(a)}<0$ these densities may differ in sign. Let us consider in more detail the laws of transformation of macroscopic scalars.

So, let further there be a $N$-plet of scalar fields:
\begin{equation}\label{n-plet}
\mathbf{\Phi}=\{\Phi_1,\Phi_2,\ldots,\Phi_N\}.
\end{equation}
Let us first find out how the macroscopic scalar densities \eqref{6a} -- \eqref{6e} are transformed with respect to the transformation of the reflection of scalar fields:
\begin{equation}\label{trans_F}
\mathbf{\Phi}:\; \mathbf{\Phi}\rightarrow -\mathbf{\Phi}.
\end{equation}
The dynamic mass of particles is transformed according to the following law at transformations $\mathbf{\Phi}$ (\ref{trans_F}):
\begin{equation}\label{trans_m}
m_{(a)}(-\mathbf{\Phi})=-m_{(a)}(\mathbf{\Phi}).
\end{equation}
It follows from the formulas \eqref{6a} -- \eqref{6e} that all macroscopic scalars can explicitly depend on scalar fields only by means of dynamic mass:
\begin{equation}\label{macr_scal(F)}
n_{(a)}=n_{(a)}(m_{(a)},\mu_{(a)}(\mathbf{\Phi}),\theta);\;\varepsilon_{(a)}=\varepsilon_{(a)}(m_{(a)}(\mathbf{\Phi}),\mu_{(a)},\theta);\;
p_{(a)}=p_{(a)}(m_{(a)}(\mathbf{\Phi}),\mu_{(a)},\theta);\; \sigma^r_{(a)}=\sigma^r_{(a)},\mu_{(a)}(\mathbf{\Phi}),\theta).
\end{equation}

Thus, we obtain the transformation laws of macroscopic scalar densities \eqref{6a} -- \eqref{6e} with respect to
to the transformation $\mathbf{\Phi}$  (\ref{trans_F}):
\begin{eqnarray}
\label{n,e}
n_{(a)}(-\mathbf{\Phi})=\overline{n}_{(a)}(\mathbf{\Phi}); & \varepsilon_{(a)}(-\mathbf{\Phi})=\overline{\varepsilon}_{(a)}(\mathbf{\Phi});\\
\label{p,s}
p_{(a)}(-\mathbf{\Phi})=\overline{p}_{(a)}(\mathbf{\Phi}); & \sigma^{r}_{(a)}(-\mathbf{\Phi})=-\overline{\sigma}^{r}_{(a)}(\mathbf{\Phi}),
\end{eqnarray}
i.e., {\it macroscopic scalar densities, except for the scalar charge density, are invariant under transformation(\ref{trans_F})}.
In this case, under the condition of continuity of the function $\sigma^r_{(a)}(\mathbf{\Phi}_r)$ due to \eqref{p,s} the following relation fulfills\footnote{see section \ref{sect_deg}}:
\begin{equation}\label{sigma(0)}
\sigma^{r}_{(a)}(\mathbf{0})=\mathbf{0}.
\end{equation}
This is an important symmetry property due to which the macroscopic equations of scalar fields with sources \eqref{macr_eqs} always have the trivial solution:
\begin{equation}\label{F=0}
\eqref{sigma(0)}\Rightarrow e_r\Box\mathbf{\Phi}_0+V'_{\mathbf{\Phi}_0}\equiv-8\pi \sigma(\mathbf{\Phi}_0),\qquad  \mathbf{\Phi}_0\equiv \mathbf{0}.
\end{equation}
Let us note that the presence of an addend with constant rest mass in the formula for the dynamic mass \eqref{m_*(pm)} would lead to the absence of a trivial solution for the macroscopic equations of scalar fields with sources.

Let us consider now the scalar charge conjugation transformation:
\begin{equation}\label{qtrans}
\mathbf{Q}:\; q^{r}_{(a)}\longleftrightarrow -q^{r}_{(a)},\Rightarrow \mathbf{q} \longleftrightarrow -\mathbf{q},
\end{equation}
where $\mathbf{q}$ -- is a charge matrix of an order $n\times N$:
\[%
\mathbf{q}=\left(\begin{array}{cccc}
q^{1}_{(1)} & q^{1}_{(2)} & \cdots & q^{1}_{(n)}\\
q^{2}_{(1)} & q^{2}_{(2)} & \cdots & q^{2}_{(n)}\\
\cdots & \cdots & \cdots & \cdots \\
q^{N}_{(1)} & q^{N}_{(2)} & \cdots & q^{N}_{(n)}\\
\end{array}\right).
\]
The transformation \eqref{qtrans} replaces scalar charged particles with antiparticles. It is easy to see that the formulas for all macroscopic scalars
\eqref{6a} -- \eqref{6e} are charge symmetric, including $\sigma^r_{(a)}$, in which the specificity of scalar interaction is revealed (see, for instance, \cite{Ignat15_2}):
\begin{eqnarray}
\label{n_E_anti}
n_{(a)}(-\mathbf{q},-\gamma)= \overline{n}_{(a)}(\mathbf{q},\gamma); &  \varepsilon_a(-\mathbf{q},-\gamma)= \overline{\varepsilon}_a(\mathbf{q},\gamma);\\
\label{p_s_anti}
p_{(a)}(-\mathbf{q},-\gamma)= \overline{p}_{(a)}(\mathbf{q},\gamma); &  \sigma^{r}_{(a)}(-\mathbf{q},-\gamma)=
\overline{\sigma}^{r}_{(a)}(\mathbf{q},\gamma).
\end{eqnarray}

Thus, the following statement is true.
\begin{stat}\label{macr_scalar_m}
The choice of the dynamic mass in the form \eqref{m_*(pm)}, which allows negative values in scalar fields, leads to the correct sign of macroscopic energy density and pressure of scalar charged particles under local thermodynamic equilibrium conditions, as in the case of microscopic density (see note \ref{note_sign}).
\end{stat}

So, under LTE conditions, the macroscopic conservation laws formally give $4+N$ independent equations on $5+N$ macroscopic scalar functions $\varepsilon_p, p_p, n_r$ and 3 independent components of the velocity vector $u^i$\footnote{one of the equations (\ref{2a}) is dependent on the rest %
due to the identity (\ref{6})}. However, not all of the indicated macroscopic scalars are functionally %
independent, since they are all determined by the locally equilibrium distribution functions \eqref{8_0}.
Under the resolved series of chemical equilibrium conditions, when only one chemical potential remains independent, the resolved mass surface equation\footnote{details see in \cite{Ignat14_1,Ignat14_2}.}, and given scalar potentials and the $4+2N$ scale factor, the macroscopic scalars  $\varepsilon_p, p_p, n_r, \sigma^r$ are determined by two scalars - certain chemical potential  $\mu$ and local temperature $\theta$. Thus, the system of equations (\ref{2a}) -- (\ref{2c}) becomes fully determined.

Let us further investigate the question of how the equations of scalar fields \eqref{macr_eqs} are transformed with respect to the transformations of the reflection of scalar fields \eqref{trans_F} and charge conjugation \eqref{qtrans}. Since the Higgs potentials \eqref{Higgs} are even functions with respect to the scalar fields $\Phi_r$, then at transformation \eqref{trans_F} it is $V'_{\Phi_r}(\mathbf{\Phi})=-V'_{\Phi_r}(-\mathbf{\Phi})$. Taking into account the transformation law of the scalar charge density \eqref{p,s}, we come to the conclusion that the equations of scalar fields with the source \eqref{macr_eqs} are invariant with respect to the transformation of reflection \eqref{trans_F}. Further, since the scalar charges $q^r_{(a)}$  are not contained in the left-hand sides of the scalar field equations \eqref{macr_eqs}, and the scalar charge densities $\sigma^r$ according to \eqref{p_s_anti} are invariant under the charge conjugation transformation \eqref{qtrans}, then the equations of scalar fields with the source \eqref{macr_eqs} are invariant with respect to the charge conjugation transformation \eqref{qtrans}. Thus, the following statement is true.

\begin{stat}\label{Inv_Eq_S}
Macroscopic equations of scalar fields with sources \eqref{macr_eqs} are invariant both with respect to the transformation of scalar fields' reflection \eqref{trans_F}, and with respect to the charge conjugation transformation \eqref{qtrans}, and in this case they always have a trivial solution \eqref{F=0}. This means, firstly, that if $\mathbf{\Phi}=\mathbf{\Phi}_0(x^i)$ is a solution for the field equations \eqref{macr_eqs}, then $\mathbf{\Phi}=\mathbf{-\Phi}_0(x^i)$ under equivalent initial conditions\footnote{with a simultaneous change of the signs of the initial scalar potentials in the Cauchy problem} and equal parameters is also a solution for these field equations. Secondly, if $\mathbf{\Phi}=\mathbf{\Phi}_0(x^i)$ is a solution of the field equations \eqref{macr_eqs} at given scalar charges $\mathbf{Q}$, then it is is a solution of these equations for the charge-conjugate problem $\mathbf{Q}\to\mathbf{-Q}$. In particular, this implies that the signs of scalar potentials do not depend on the signs of the charges of their sources and are determined only by the initial conditions of the Cauchy problem.
\end{stat}

These features of the interaction of scalar fields with scalar charges radically distinguish this interaction from th e interaction of vector and tensor fields with the corresponding charges. We can say that in the presence of scalar fields, a system of particles is charged with a scalar charge. Situation's peculiarity is revealed in the fact that scalar charges are practically conserved according to the charge conservation law \eqref{2c}, but their effective densities, which are the sources of scalar fields,  $\sigma^r$, fully depend on the value of scalar fields $\Phi_r$ by means of dynamic mass.

\subsection{Locally Equilibrium Statistical System of Scalar Charged Fermions Under Conditions of Complete Degeneracy \label{sect_deg}}
Further in this article we will consider the equilibrium statistical system of fermions under conditions of complete degeneracy. In the case of complete degeneracy of the statistical system of fermions it is
\begin{equation}\label{deg}
\theta\to 0,
\end{equation}
the locally equilibrium distribution function of fermions \eqref{8_0} takes the form of a step function \cite{Ignat14_2}:
\begin{equation}\label{f_deg}
f_{(a)}^0(x,P)=\chi_+\left(\mu_{(a)}-\sqrt{m_{(a)}^2+P^2}\right),
\end{equation}
where $\chi_+(z)$ is the Heaviside step function, $\mu_{(a)}\equiv \sqrt{m_{(a)}^2+f_{(a)}^2}$, $f_{(a)}$ -- is the Fermi momentum for the ``a''-sort of fermions.

The result of integration of macroscopic densities \eqref{6a} -- \eqref{6e} with respect to the distribution \eqref{f_deg}
is expressed in elementary functions  \cite{Ignat14_2}. Thus, for a multicomponent Fermi system with a multi-particle charge, we obtain the following expressions for macroscopic scalars:
\begin{equation}\label{2_3}
n^{(a)}=\frac{1}{\pi^2}\pi_{(a)}^3;
\end{equation}
\begin{equation}\label{2_3a}
{\displaystyle
\begin{array}{l}
\varepsilon_p = {\displaystyle\sum\limits_a\frac{m_{(a)}^4}{8\pi^2}}F_2(\psi_{(a)});
\end{array}}
\end{equation}
\begin{equation}\label{2_3b}{\displaystyle
\begin{array}{l}
p_p ={\displaystyle\sum\limits_a\frac{m_{(a)}^4}{24\pi^2}}(F_2(\psi_{(a)})-4F_1(\psi_{(a)}))
\end{array}}
\end{equation}
\begin{equation}\label{2_3c}{\displaystyle
\begin{array}{l}
\sigma^r={\displaystyle \sum\limits_a q^r_{(a)}\frac{m_{(a)}^3}{2\pi^2}}F_1(\psi_{(a)}),
\end{array}}
\end{equation}
where $m_{(a)}$ is described by formula \eqref{m_*(pm)} and dimensionless functions $\psi_{(a)}$ are introduced
\begin{equation}\label{psi}
\psi_{(a)}=\frac{\pi_{(a)}}{|m_{(a)}|},
\end{equation}
which are equal to relation of the Fermi momentum $\pi_{(a)}$ to the total energy of a fermion, and to reduce the notation the functions  $F_1(\psi)$ and $F_2(\psi)$ are introduced:
\begin{equation}\label{F_1}
F_1(\psi)=\psi\sqrt{1+\psi^2}-\ln(\psi+\sqrt{1+\psi^2});
\end{equation}
\begin{equation}
\label{F_2}
F_2(\psi)=\psi\sqrt{1+\psi^2}(1+2\psi^2)-\ln(\psi+\sqrt{1+\psi^2}).
\end{equation}
Functions $F_1(x)$ and $F_2(x)$ are first of all, even:
\begin{equation}\label{F(-x)}
F_1(-x)=-F_1(x);\quad F_2(-x)=-F_2(x),
\end{equation}
and secondly, have the following asymptotics
\begin{eqnarray}\label{F,x->0}
\left.F_1(x)\right|_{x\to 0}\simeq \frac{2}{3}x^3; \left.F_2(x)\right|_{x\to 0}\simeq \frac{8}{3}x^3;\nonumber\\
\left.(F_2(x)-4F_1(x))\right|_{x\to 0}\simeq \frac{8}{5}x^5;
\end{eqnarray}
\begin{eqnarray}\label{F,x->8}
\left.F_1(x)\right|_{x\to\pm\infty}\simeq x|x|;\quad \left.F_2(x)\right|_{x\to\pm\infty}\simeq 2x^3|x|.
\end{eqnarray}
Let us also write down expressions for the derivatives of the functions $F_1(x)$ and $F_2(x)$ that are useful further:
\begin{eqnarray}\label{F'_12}
F'_1(x)=\frac{2x^2}{\sqrt{1+x^2}};\quad F'_2(x)=8x^2\sqrt{1+x^2}.
\end{eqnarray}

Summing up both parts of the relations \eqref{2_3a} and \eqref{2_3b}, one can easily verify the validity of the identity:
\begin{equation}\label{E_P_f}
\varepsilon_p+p_p\equiv \frac{1}{3\pi^2}\sum\limits_a m^4_{(a)}\psi_{(a)}^3\sqrt{1+\psi_{(a)}^2}.
\end{equation}

To conclude this section, we find the asymptotics for the macroscopic scalars \eqref{2_3} -- \eqref{2_3c} at $m_{(a)}\to 0$, i.e., at  $\mathbf{\Phi}\to0$. It is necessary to take into account, first, the formula \eqref{psi} and, secondly, the asymptotic formulas for the functions $F_1(x),F_2(x)$ \eqref{F,x->8}. Thus, we find:
\begin{eqnarray}\label{macr_scal-m->0}
\left.n_{(a)}\right|_{\mathbf{\Phi}\to0}= \frac{1}{\pi^2}\pi_{(a)}^3=\mathrm{Const};\quad
\left.\sigma^r_{(a)}\right|_{\mathbf{\Phi}\to0}\simeq q^r_{(a)}m_{(a)}\frac{\pi^2_{(a)}}{2\pi^2}\to 0;\nonumber\\
\left.\varepsilon_{(a)}\right|_{\mathbf{\Phi}\to0}\simeq 3\left.p_{(a)}\right|_{\mathbf{\Phi}\to0}\simeq \frac{1}{4\pi^2}\sum\limits_a \pi^4_{(a)}=\mathrm{Const}.
\end{eqnarray}
Thus, as we indicated above, $\sigma^r_{(a)}(\mathbf{\mathbf{\Phi}})$ is a continuous function for $\mathbf{\Phi}\to\mathbf{0}$, and in accordance with \eqref{sigma(0)}, $\sigma^r_{(a)}(\mathbf{0})=\mathbf{0}$.

\section{The Cosmological Model}
We will consider a homogeneous isotropic distribution of matter, assuming that space-time is described by the spatially flat Friedmann metric
\begin{equation}\label{ds_0}
ds^2=dt^2-a^2(t)(dx^2+dy^2+dz^2),
\end{equation}
and all thermodynamic functions and scalar fields depend only on time. It is easy to verify that $u^i=\delta^i_4$ turns equations (\ref{2a}) into identities, and the system of equations (\ref{2b}) -- (\ref{2c}) is reduced to $1+N$ \emph{material} equations\footnote{Here we write down the material equations for the case of $N$ scalar fields.}:
\begin{equation}\label{7a1}
\dot{\varepsilon}_p+3\frac{\dot{a}}{a}(\varepsilon_p+p_p)=\sum\limits_r\sigma^r\dot{\Phi}_r;
\end{equation}
\begin{equation}\label{7b1}
\dot{\rho}^r +3\frac{\dot{a}}{a}\rho^r=0 \quad(r=\overline{1,N}),
\end{equation}
where $\dot{\phi}\equiv d\phi/dt$.

\subsection{Solution of the Material Equations for a Degenerate Plasma }
Thus, there remain $N+1$  differential equations for $n+1$ thermodynamic functions $\mu_{(a)}$ and $\theta$. Proceeding to the limit $\mu_{(a)}\to \infty$ (or $\theta\to0$) we obtain a system of $N+1$ equations on $т$ functions, and the problem of compatibility of these equations arises, and this the problem is independent of the presence of a scalar field. However \cite{Ignat14_2} shows
that this is an apparent problem, in fact, no contradictions in the system of equations, (\ref{2a}) -- (\ref{2c}) arise even in the case of a degenerate Fermi system. As it turns out, in this case, the laws of conservation of charge \eqref{7b1} are a direct consequence of the law of conservation of energy \eqref{7a1}.

Indeed, let us consider the energy conservation law \eqref{7a1} or the degenerate Fermi-system. Using the expressions \eqref{2_3a} -- \eqref{2_3c} and \eqref{E_P_f} for macroscopic scalars,  definitions of the dynamic mass of particles $m_{(a)}$ \eqref{m_*(pm)}, functions $\psi_{(a)}$ \eqref{psi}, $F_1(x)$ and $F_2(x)$ \eqref{F_1} and \eqref{F_2}, as well as expressions for their derivatives \eqref{F'_12}, after rather cumbersome calculations, we bring the energy conservation law \eqref{7a1} to the explicit form:
\begin{equation}\label{cons_E_p}
\sum\limits_{a=1}^n m_{(a)}\pi_{(a)}^3\sqrt{1+\psi_{(a)}^2}\frac{d}{dt}\ln(|a \pi_{(a)}|)=0,
\end{equation}
which, with an account of the definition of the total mass \eqref{m_*(pm)} can be represented in the explicit form:
\begin{equation}\label{cons_E_p1}
\sum\limits_{a=1}^n \pi_{(a)}^3\sqrt{1+\psi_{(a)}^2}\frac{d}{dt}\ln(|a \pi_{(a)}|)\sum\limits_{r=1}^N q^r_{(a)}\Phi_r=0.
\end{equation}
Further, taking into account the definition of the kinematic charge density \eqref{n^r} and the expressions for the density of the number of particles \eqref{2_3} we reduce the conservation laws of scalar charges \eqref{7b1} to the form:
\begin{equation}\label{cons_n}
\sum\limits_{a=1}^n \pi_{(a)}^3 \frac{d}{dt}\ln(|a \pi_{(a)}|)\ q^r_{(a)} =0, \quad (r=\overline{1,N}).
\end{equation}

The equations \eqref{cons_E_p1} -- \eqref{cons_n} are a system of $N+1$ homogeneous algebraic equations in $n$ unknowns
$X_{(a)}=d/dt(a\pi_{(a)})$, $a=\overline{1,n}$. This system always has a trivial solution $X_{(a)}=0$, and for $n\leqslant N+1$ in the general case it has only the trivial solution.  We will choose this solution as a solution of the material equations \eqref{7a1} and \eqref{7b1}. Let us note that for a one-component or two-component system of fermions, the solution of the material equations with respect to $X_{(a)}$ can only be trivial. However, even for any numbers $n$ and $N$ the system of equations \eqref{cons_E_p1} -- \eqref{cons_n} has only a trivial solution due to the arbitrariness of the parameters of scalar charges $q^r_{(a)}$. Thus, we choose the solution:
\begin{equation}\label{ap}
\frac{d}{dt}\ln(a \pi_{(a)})=0\Leftrightarrow a\pi_{(a)}=\mathrm{Const}.
\end{equation}
Let us note that, on the one hand, this solution corresponds to the conservation law for the absolute value of the momentum of free particles in a homogeneous isotropic metric, and on the other hand, it was obtained in \cite{Ignat17} for a one-component Fermi system.

\begin{stat}\label{stat_ap}
The solution of the material equations \eqref{7a1} -- \eqref{7b1} is the Fermi momentum conservation law \eqref{ap} for each component of the Fermi system.
\end{stat}

\subsection{Field Equations}
Further we consider the asymmetric scalar Higgs doublet $\{\Phi,\varphi\}$ \footnote{The term ``quintom'' is also used in the literature} with the Lagrange function as a field model \cite{YuI_Kokh19_1}, \cite{Ignat21_TMP}:
\begin{eqnarray}\label{Ls}
\mathrm{L}_s=\frac{1}{8\pi}\biggl(\frac{1}{2}g^{ik}\Phi_{_,i}\Phi_{_,k}-V(\Phi)\biggr)+
\frac{1}{8\pi}\biggl(-\frac{1}{2}g^{ik}\varphi_{,i}\varphi_,k-\mathcal{V}(\varphi)\biggr),
\end{eqnarray}
where $V(\Phi)$ -- is a potential energy of the classical scalar field and  $\mathcal{V}(\varphi)$ -- is a potential energy of the phantom one:
\begin{equation}\label{Higgs}
V=-\frac{\alpha}{4} \left(\Phi^{2} -\frac{m^2}{\alpha}\right)^{2},\; \mathcal{V}=-\frac{\beta}{4} \left(\varphi^{2} -\frac{\mathfrak{m}^2}{\beta}\right)^{2},
\end{equation}
$\alpha,\ \beta$ -- are constants of self-action, $m,\ \mathfrak{m}$ -- are masses of scalar bosons.

Further %
\begin{eqnarray}\label{T_{iks}}
T^{ik}_s=\frac{1}{8\pi}\biggl(\Phi^{,i}\Phi^{,k}-
\frac{1}{2}g^{ik}\Phi_{,j}\Phi^{,j}+g^{ik} V(\Phi)\biggr)+\frac{1}{8\pi}\biggl(-\varphi^{,i}\varphi^{,k}+
\frac{1}{2}g^{ik}\varphi_{,j}\varphi^{,j}+g^{ik} \mathcal{V}(\varphi)\biggr)
\end{eqnarray}
-- is a tensor of scalar doublet's energy - momentum. Let us note that we can omit the constant terms in the Higgs potentials \eqref{Higgs}, redefining the cosmological constant $\Lambda_0$
\begin{equation}\label{L->L}
\Lambda\to \Lambda_0-\frac{m^4}{4\alpha}-\frac{\mathfrak{m}^4}{4\beta}.
\end{equation}
Scalar fields are defined by equations for charged scalar fields with a source \cite{Ignat15}:
\begin{equation}\label{Eq_Phi-varphi}
\square\Phi+V'_\Phi=-8\pi\sigma_c,\qquad -\square\varphi+v'_\varphi=-8\pi\sigma_f,
\end{equation}
where $\sigma_c,\sigma_f$ -- are scalar charge densities \eqref{2_3c} for classical and phantom fields, correspondingly. It can be shown that due to \eqref{2} and \eqref{Eq_Phi-varphi} the conservation law of the total energy-momentum tensor of the ``plasma + charged scalar fields'' system is identically fulfilled:
\begin{equation}\label{dTik=0}
\nabla_iT^{ik}=\nabla_i\bigl(T^{ik}_{p}+T^{ik}_s\bigr)\equiv0.
\end{equation}

The energy-momentum tensor of an asymmetric scalar doublet in an isotropic homogeneous metric also takes the form of an energy-momentum tensor of an ideal isotropic fluid:
\begin{equation} \label{MET_s}
T_{s}^{ik} =(\varepsilon_s +p_{s} )u^{i} u^{k} -p_s g^{ik} ,
\end{equation}
where it is:
\begin{eqnarray}\label{Es}
\varepsilon_s=\frac{1}{8\pi}\biggl(\frac{1}{2}\dot{\Phi}^2+V(\Phi)-\frac{1}{2}\dot{\varphi}^2+\mathcal{V}(\varphi)\biggr);\\
\label{Ps}
p_{s}=\frac{1}{8\pi}\biggl(\frac{1}{2}\dot{\Phi}^2-V(\Phi)-\frac{1}{2}\dot{\varphi}^2-\mathcal{V}(\varphi)\biggr),
\end{eqnarray}
so that:
\begin{equation}\label{e+p}
\varepsilon_s+p_{s}=\frac{1}{4\pi}\bigl(\dot{\Phi}^2-\dot{\varphi}^2\bigr).
\end{equation}
Finally, the equations of scalar fields \eqref{Eq_Phi-varphi} in the Friedman metric take the form:
\begin{eqnarray}\label{Eq_S_t1}
\ddot{\Phi}+3\frac{\dot{a}}{a}\dot{\Phi}+m^2\Phi-\alpha\Phi^3= -8\pi\sigma_c(t),\\
\label{Eq_S_t2}
\ddot{\varphi}+3\frac{\dot{a}}{a}\dot{\varphi}-\mathfrak{m}^2\varphi+\beta\varphi^3= 8\pi\sigma_f(t).
\end{eqnarray}
We will further consider the standard Einstein equations with an \emph{unperturbed} $\Lambda$-term, $\Lambda_0$:
\begin{equation}\label{EqEinst}
G^i_k\equiv R^i_k-\frac{1}{2}R\delta^i_k=8\pi T^i_k+\Lambda_0\delta^i_k.
\end{equation}
Let us write down the independent Einstein equations with respect to the Friedmann metric \eqref{ds_0}:
\begin{equation}\label{Einstein^1_1}
2\frac{\ddot{a}}{a}+\frac{\dot{a}^2}{a^2}+\sum\limits_r\biggl(\frac{e_r\dot{\Phi_r^2}}{2}
-\frac{m_r^2\Phi_r^2}{2}+\frac{\alpha_r\Phi_r^4}{4}\biggr)+8\pi p_p=\Lambda;
\end{equation}
\begin{equation}\label{Einstein^4_4}
3\frac{\dot{a}^2}{a^2}-\sum\limits_r\biggl(\frac{e_r\dot{\Phi_r^2}}{2}
+\frac{m_r^2\Phi_r^2}{2}-\frac{\alpha_r\Phi_r^4}{4}\biggr)-8\pi\varepsilon_p=\Lambda,
\end{equation}
where $\Lambda$ -- is a renormalized cosmological constant \eqref{L->L}.

one can obtain an independent Einstein equation \cite{Ignat20} from \eqref{Einstein^1_1} and \eqref{Einstein^4_4}:
\begin{equation}\label{(11-44)-0}
\dot{H}+4\pi(\varepsilon+p)=0,
\end{equation}
where $\varepsilon=\varepsilon_p+\varepsilon_s$ and $p=p_p+p_s$. Next, after \cite{Ignat20} we introduce the total energy, $\mathcal{E}$, of the cosmological matter:
\begin{equation}\label{E}
\mathcal{E}=\frac{1}{8\pi}(3H^2-\Lambda_0)-\varepsilon,
\end{equation}
with a help of which, the Einstein equation \eqref{Einstein^4_4} can be given a simple form:
\begin{equation}\label{E=0}
\mathcal{E}=0,
\end{equation}
which reflects the fact that the total energy of a space - flat Friedmann Universe is equal to zero.

Differentiating the total energy over time \eqref{E} and taking into account the field equations \eqref{Eq_S_t1} -- \eqref{Eq_S_t2} and the relations \eqref{7a1}, \eqref{Es}, \eqref{e+p} and \eqref{(11-44)-0}, we obtain the energy conservation law
\begin{equation}\label{dE/dt=0}
\frac{d}{dt}\mathcal{E}=0\Rightarrow \mathcal{E}=\mathcal{E}_0.
\end{equation}
Thus, the consequence of the considered system of dynamical equations \eqref{7a1}, \eqref{Eq_S_t1} -- \eqref{Eq_S_t2} и \eqref{(11-44)-0} is the conservation law for the total energy of the cosmological system \eqref{dE/dt=0} $\mathcal{E}=\mathcal{E}_0$. Here the Einstein equation \eqref{E=0}
is the partial integral of this system $\mathcal{E}_0=0$. A similar situation arises for vacuum scalar fields \cite{Ignat20}. This means that the first integral \eqref{E=0} can be considered as an initial condition in the Cauchy problem for the cosmological model.

In particular, for a degenerate Fermi system, taking into account \eqref{E_P_f} we find the following equation from \eqref{(11-44)-0}:
\begin{equation}\label{11-44}
\dot{H}+\frac{\dot{\Phi}^2}{2}-\frac{\dot{\varphi}^2}{2}+\frac{4}{3\pi}\sum\limits_a m^4_{(a)}\psi_{(a)}^3\sqrt{1+\psi_{(a)}^2}=0.
\end{equation}

\begin{stat}\label{stat_full_sys}
The system of equations \eqref{7a1} with its found solution \eqref{psi}, \eqref{Eq_S_t1}, \eqref{Eq_S_t2} and \eqref{11-44} together with definitions \eqref{2_3a} -- \eqref{2_3c} and the integral condition \eqref{Einstein^4_4} describes a closed mathematical model of the cosmological evolution of a completely degenerate Fermi system with interaction with an asymmetric Higgs scalar doublet.
\end{stat}

Following \cite{Ignat20} we also introduce the nonnegative, according to  \eqref{E} and \eqref{E=0} effective energy of the cosmological system $\mathcal{E}_{eff}$
\begin{eqnarray}\label{E_eff}
\mathcal{E}_{eff}=\varepsilon+\frac{\Lambda_0}{8\pi}\geqslant 0\Rightarrow \mathcal{E}_{eff}=\frac{\dot\Phi^2}{2}-\frac{\dot{\varphi}^2}{2}+V(\Phi)+\mathcal{V}(\varphi)+\frac{\Lambda_0}{8\pi}\geqslant 0,
\end{eqnarray}
as well as an explicit expression for the invariant cosmological acceleration \eqref{H=,Omega=} (see, e.g., \cite{Ignat20})
\begin{equation}\label{Omega(H)}
\Omega=1+\frac{\dot{H}}{H^2}.
\end{equation}
\subsection{Autonomous Normal System of Equations}
Choosing the variable $\xi(t)$ instead of the nonnegative dynamic variable $a(t)\geqslant0$
\begin{eqnarray}\label{xi(t)}
\xi=\ln a,\quad \xi\in(-\infty,+\infty);&\\
\label{dxi/dt}
\dot{\xi}=H,&
\end{eqnarray}
let us use the integral \eqref{ap} of the conservation law of plasma \eqref{cons_E_p} for finding the function $\psi_{(a)}(t)$ \eqref{psi}
\begin{equation}\label{psi0}
\psi_{(a)}=\frac{\pi^0_{(a)} \mathrm{e}^{-\xi}}{|m_{(a)}|}, \quad (\pi^0_{(a)}=\pi_{(a)}(0)).
\end{equation}
Thus, taking into account \eqref{m_*(pm)} we obtain the expression for the function $\psi(t)$:
\begin{equation}\label{psi1}
\psi_{(a)}=\frac{\pi^0_{(a)}}{\bigl|\sum_r q^r_{(a)}\Phi_r\bigr|}\mathrm{e}^{-\xi}
\end{equation}
and scalar charge density \eqref{2_3c}
\begin{equation}\label{sigma1}
\sigma^r=\frac{1}{2\pi^2}\sum\limits_a q^r_{a}F_1(\psi_{a)})\biggl(\sum_{r'}q^{r'}_{(a)}\Phi_{r'}\biggr)^3.
\end{equation}
Assuming further for the model of an asymmetric scalar doublet
\begin{equation}\label{dPhi/dt}
\dot{\Phi}=Z, \quad \dot{\varphi}=z,
\end{equation}
we obtain in this notation the dynamic equations that represent a normal autonomous system of 6 ordinary differential equations \eqref{dxi/dt}, \eqref{dPhi/dt}, \eqref{11-44_Zz}, \eqref{Eq_SZ_t1} and \eqref{Eq_Sz_t2} with respect to 6 dynamic variables $\xi,H$, $\Phi,Z$, $\varphi,z$
\begin{eqnarray}\label{11-44_Zz}
\dot{H}=-\frac{Z^2}{2}+\frac{z^2}{2}-\frac{4}{3\pi}\sum\limits_a m^4_{(a)}\psi_{(a)}^3\sqrt{1+\psi_{(a)}^2}=0;\\
\label{Eq_SZ_t1}
\dot{Z}=-3\frac{\dot{a}}{a}Z-m^2\Phi+\alpha\Phi^3-8\pi\sigma_c,\\
\label{Eq_Sz_t2}
\dot{z}=-3\frac{\dot{a}}{a}z+\mathfrak{m}^2\varphi-\beta\varphi^3+ 8\pi\sigma_f,
\end{eqnarray}
where expression for $\sigma_c$ is obtained from \eqref{sigma1} by summation in outer sum over classical charges $q^c_{(a)}$, and the expression for
$\sigma_f$ -- by summation over phantom charges $q^f_{(a)}$.

In these variables, the first integral \eqref{Einstein^4_4} takes the form
\begin{equation}\label{Surf_Einst}
\Sigma_E:\; 3H^2-\Lambda-\frac{Z^2}{2}+\frac{z^2}{2}+\frac{m^2}{2}\Phi^2-\frac{\alpha\Phi^4}{4}+\frac{\mathfrak{m}^2}{2}\varphi^2-\frac{\beta\varphi^4}{4}
-\frac{1}{\pi}\sum\limits_a m^4_{(a)}F_2(\psi_{(a)})=0,
\end{equation}
The \eqref{Surf_Einst} equation is an algebraic equation with respect to dynamical variables $\xi,H$, $\Phi,Z$, $\varphi,z$ and describes a certain 5-dimensional hypersurface $\Sigma_E$ in a 6-dimensional arithmetic phase space of the dynamical system $\mathbb{R}_6=\{\xi,H,\Phi,Z,\varphi,z\}$. Following \cite{Ignat20} this hypersurface will be called the \emph{Einstein hypersurface}. All phase trajectories of the dynamical system \eqref{dxi/dt}, \eqref{dPhi/dt}, \eqref{11-44_Zz}, \eqref{Eq_SZ_t1} and \eqref{Eq_Sz_t2}, like the initial points, must lie on the hypersurface Einstein. Since \eqref{Surf_Einst1} is the first integral of the dynamical system, to solve the Cauchy problem it is sufficient to require that the initial point of the dynamical trajectory belong to the Einstein hypersurface (see \cite{Ignat20}, \cite{Ignat21_TMP}). Thus, the following statement is true.

\begin{stat}\label{stat_dyn_sys}
The dynamical system of the cosmological model, based on a locally equilibrium system of scalar charged degenerate fermions, is fully described by the normal autonomous system of 6 ordinary differential equations \eqref{dxi/dt}, \eqref{dPhi/dt}, \eqref{11-44_Zz}, \eqref{Eq_SZ_t1} and \eqref{Eq_Sz_t2} with a single algebraic condition \eqref{Surf_Einst}, which is the partial value of the first integral of this system. All trajectories of the dynamical system in the phase space $\mathbb{R}_6=\{\xi,H,\Phi,Z,\varphi,z\}$ lie on the Einstein hypersurface $\Sigma_E$ \eqref{Surf_Einst}. Due to the autonomy of the dynamical system, it is invariant with respect to time translations
\begin{equation}\label{t->t}
t\to t+t_0,
\end{equation}
which, in particular, allows us to always choose $t_0$ so that the condition is fulfilled
\begin{equation}\label{a(0)}
a(0)=1\Rightarrow \xi(0)=0.
\end{equation}
\end{stat}

\subsection{Two Simple Models of a Fermion System}
Below in this article we will consider two fundamentally different simplest models of the statistical system of scalar charged fermions with an asymmetric scalar doublet.
\\[12pt]
\textbf{Model 1}. A two-component statistical system of degenerate fermions, where the particles of one component have the classical charge $e$ and the particles of the second component have a phantom scalar charge $\epsilon$. In this case, two terms remain under the summation sign on the right-hand sides of the formulas \eqref{6a} -- \eqref{6e}. We will label the masses of fermions - carriers of the corresponding charges with $m_e$ and $m_\epsilon$:
\begin{equation}\label{mF-mf}
m_e=e\Phi,\quad m_\epsilon=\epsilon\varphi.
\end{equation}
\textbf{Model 2}.  One-component statistical system of fermions, scalar charged with two charges $\{e,\epsilon\}$ under conditions of complete degeneration.
In this case, only one term remains under the summation sign on the right-hand sides of the formulas \eqref{6a} -- \eqref{6e}. For the mass of the charge in this case, we have the formula:
 \begin{equation}\label{mFf}
 m=e\Phi+\epsilon\varphi.
 \end{equation}

The difference between the models \textbf{1} and \textbf{2} is concluded, firstly, in the difference in formulas for dynamic masses - the first model has two dynamic masses $m_c=e\Phi$ and $m_f=\epsilon\varphi$, while in the second, there is only one $m=e\Phi+\epsilon\varphi$. Secondly, in the second model there are 2 Fermi momenta $\pi_c$ and $\pi_f$ accordingly, there are two functions $\psi$: $\psi_c=\pi_c/m_c(\Phi)$ and $\psi_f=\pi_f/m_f(\varphi)$. Finally, in the corresponding formulas for macroscopic scalars, it is required to carry out summation over the two components of the statistical system.

Further, in the case of the model \textbf{1} it is required to carry out summation over 2 fermion components in the equation \eqref{11-44}  and in the case of model \textbf{2} -- just to keep a single term under the sign of the sum.

\section{The Analysis of Model 1 with a Two-Component Fermi System of Singly Scalarly Charged Fermions}
\subsection{The Dynamic System for the Model 1}
In this case, the formulas \eqref{psi1} and \eqref{sigma1} take the form:
\begin{equation}\label{psi-psi}
\psi_c=\frac{\pi^0_c}{|e\Phi|}\mathrm{e}^{-\xi}; \quad \psi_f=\frac{\pi^0_f}{|\epsilon\varphi|}\mathrm{e}^{-\xi};
\end{equation}
\begin{equation}\label{sigma-sigma}
\sigma_c=\frac{e^4}{2\pi^2}\Phi^3 F_1(\psi_c); \quad \sigma_f=\frac{\epsilon^4}{2\pi^2}\varphi^3 F_1(\psi_f).
\end{equation}
Further, the equations of the scalar doublet filds \eqref{Eq_S_t1} -- \eqref{Eq_S_t2} with an account of \eqref{dPhi/dt} and \eqref{sigma-sigma} can be written in the following form:
\begin{eqnarray}\label{dZ/dt}
\dot{Z}=-3HZ-m^2\Phi +\Phi^3\biggl(\alpha-\frac{4e^4}{\pi}F_1(\psi_c)\biggr);\\
\label{dz/dt}
\dot{z}=-3Hz+\mathfrak{m}^2\varphi -\varphi^3\biggl(\beta-\frac{4\epsilon^4}{\pi}F_1(\psi_f)\biggr).
\end{eqnarray}
Einstein's equation  \eqref{11-44} and the first integral of the system of equations \eqref{Surf_Einst} or model 1 takes the form:
\begin{eqnarray}\label{dH/dt_0}
\dot{H}=- \frac{Z^2}{2}+ \frac{z^2}{2}-\frac{4}{3\pi}e^4\Phi^4\psi_c^3\sqrt{1+\psi_c^2}-
\frac{4}{3\pi}\epsilon^4\varphi^4\psi_f^3\sqrt{1+\psi_f^2};
\end{eqnarray}
\begin{eqnarray}\label{Surf_Einst1}
3H^2-\Lambda-\frac{Z^2}{2}+\frac{z^2}{2}
-\frac{m^2\Phi^2}{2}+\frac{\alpha\Phi^4}{4}-\frac{\mathfrak{m}^2\varphi^2}{2}+\frac{\beta\varphi^4}{4}-\frac{e^4\Phi^4}{\pi}F_2(\psi_c)-\frac{\epsilon^4\varphi^4}{\pi}F_2(\psi_f)=0.
\end{eqnarray}

Further, the points of the phase space $\mathbb{R}_4$, the effective energy \eqref{E_eff} s negative, are inaccessible to the dynamical system.
The inaccessible region is separated from the accessible region of the phase space by the zero effective energy hypersurface $S_{E}\subset \mathbb{R}_6$, which is a cylinder with the $OH$ axis:
\begin{eqnarray}\label{S_E1}
S_E:\; \Lambda+\frac{e^4\Phi^4}{\pi}F_2(\psi_c)+\frac{\epsilon^4\varphi^4}{\pi}F_2(\psi_f)
+\frac{Z^2}{2}-\frac{z^2}{2}+\frac{m^2\Phi^2}{2}-\frac{\alpha\Phi^4}{4}+\frac{\mathfrak{m}^2\varphi^2}{2}-\frac{\beta\varphi^4}{4}=0,
\end{eqnarray}
and the hypersurface of null effective energy \eqref{S_E1} touches the Eisntein hypersurface \eqref{Surf_Einst1} in the hyperplane
$H=0$:
\begin{equation}\label{H=0}
\Sigma_E \cap S_E =H=0.
\end{equation}

As can be seen from the equation \eqref{dH/dt_0} in the case of a scalarly neutral statistical system ($e=\epsilon\equiv0$) the sign of the derivative of the Hubble constant is completely determined by the difference $z^2-Z^2$: in the case of dominance of the classical scalar field ($Z^2>z^2$) it is always $\dot{H}<0$, and in the case of dominance of a phantom scalar field ($Z^2<z^2$) it is always $\dot{H}>0$. A variation of these factors in the course of cosmological evolution can lead to a wide variety of types of behavior of the cosmological model \cite{Ignat21_TMP}. In the presence of charged matter, its contribution to this variation, as seen from \eqref{dH/dt_0}, contributes to a decrease of the Hubble constant.

Let us note that instead of the equation \eqref{dH/dt_0} we can consider an equivalent equation (see \cite{Ignat20}), by substituting the expression for $z^2/2-Z^2/2$ from \eqref{Surf_Einst1} into \eqref{dH/dt_0}:
\begin{eqnarray}\label{dH/dt}
\dot{H}=-3H^2+\Lambda+\frac{e^4\Phi^4}{\pi}F_2(\psi_c)+\frac{\epsilon^4\varphi^4}{\pi}F_2(\psi_f)
-\frac{m^2\Phi^2}{2}-\frac{\mathfrak{m}^2\varphi^2}{2}  \nonumber\\
+\frac{\alpha\Phi^4}{4}+\frac{\beta\varphi^4}{4}+
\frac{4}{3\pi}e^4\Phi^4\psi_c^3\sqrt{1+\psi_c^2}+\frac{4}{3\pi}\epsilon^4\varphi^4\psi_f^3\sqrt{1+\psi_f^2}.
\end{eqnarray}

\subsection{Singular Points of the Dynamic System of Model 1}
Singular points of a dynamical system, represented by a normal autonomous system of differential equations, are determined by algebraic equations obtained by equating the derivatives of all dynamical variables to zero. Thus, we obtain the system of algebraic equations for finding coordinates of singular points from \eqref{dxi/dt}, \eqref{dPhi/dt},  \eqref{dZ/dt}, \eqref{dz/dt} and \eqref{dH/dt_0}:
\begin{equation}
\label{Zz=0}
Z=0;\quad z=0;
\end{equation}
\begin{equation}\label{H=0}
H=0;
\end{equation}
\begin{equation}
\label{-3HZ}
\Phi^3\biggl(\alpha-\frac{4e^4}{\pi}F_1(\psi_c)\biggr)-m^2\Phi=0;
\end{equation}
\begin{equation}\label{-3Hz}
\varphi^3\biggl(\beta-\frac{4\epsilon^4}{\pi}F_1(\psi_f)\biggr)-\mathfrak{m}^2\varphi=0;
\end{equation}
\begin{equation} \label{dotH=0}
e^4\Phi^4\psi_c^3\sqrt{1+\psi_c^2} +\epsilon^4\varphi^4\psi_f^3\sqrt{1+\psi_f^2}=0.
\end{equation}

In addition, it is necessary to take into account the integral of the total energy \eqref{Surf_Einst1}, -- the coordinates of the singular point must satisfy this equation, which, taking into account \eqref{Zz=0} takes the form:
\begin{eqnarray}\label{Surf_00}
\Lambda+\frac{e^4\Phi^4}{\pi}F_2(\psi_e)+\frac{m^2\Phi^2}{2}-\frac{\alpha\Phi^4}{4}+\frac{\epsilon^4\varphi^4}{\pi}F_2(\psi_\epsilon)+
\frac{\mathfrak{m}^2\varphi^2}{2}-\frac{\beta\varphi^4}{4}=0.
\end{eqnarray}
Let us consider first the equation \eqref{dotH=0}. Taking into account the definition of the functions $\psi_{(a)}$ \eqref{psi-psi}, we rewrite this equation in an equivalent form:
\begin{equation}\label{dotH=0_a}
\mathrm{e}^{-3\xi}\biggl(|e\Phi|(\pi^0_c)^3\sqrt{1+\psi^2_c}+|\epsilon\varphi|(\pi^0_f)^3\sqrt{1+\psi^2_f}\biggr)=0.
\end{equation}
It follows from the \eqref{dotH=0_a} that it is either 1). $\Phi=\varphi=0$, or 2). $\xi=+\infty$ (here it is $\psi_{(a)}=0$); $F_1(\psi_{(a)})=0$; $F_2(\psi_{(a)})=0$.

The first of these possibilities, given $\Lambda=0$ provides a line of singular points
\begin{equation}\label{M-xi}
M_\xi=(\xi,0,0,0,0,0)\quad (\mbox{при условии }\Lambda=0),
\end{equation}
since the zero solution for the potentials of scalar fields converts the equations \eqref{-3HZ} and \eqref{-3Hz} into identities, while the condition $\Lambda=0$ is turned to an identity by the equation \eqref{Surf_00}.

In the case of the second possibility \eqref{dotH=0} is turned to an identity, and equations \eqref{-3HZ} -- \eqref{-3Hz} except from the case of the studied above trivial solution $\Phi=\varphi=0$ leading to a straight line of singular points and condition $\Lambda=0$, have 8 more solutions at special values of the cosmological constant $\Lambda_0$:
\begin{equation}\label{M_pm_pm}
M_{\infty,\pm,\pm}=\biggl(+\infty,0,\pm\frac{m}{\sqrt{\alpha}},0,\pm\frac{\mathfrak{m}}{\sqrt{\beta},0}\biggr),\; \Lambda_0=0;
\end{equation}
\begin{equation}\label{M_pm_0}
M_{\infty,\pm,0}=\biggl(+\infty,0,\pm\frac{m}{\sqrt{\alpha}},0,0,0\biggr),\; \Lambda_0=\frac{\mathfrak{m}^4}{4\beta};
\end{equation}
\begin{equation}\label{M_0_pm}
M_{\infty,0,\pm}=\biggl(+\infty,0,0,0,\pm\frac{\mathfrak{m}}{\sqrt{\beta},0}\biggr),\; \Lambda_0=\frac{m^4}{4\alpha};
\end{equation}
where signs $\pm$ in two terms \eqref{M_pm_pm} take independent values. Let us note that the value  $\xi=+\infty$ is corresponded by $a=\infty$, i.e.,
infinite future, and value $\xi=-\infty$ is corresponded by $a=0$, i.e., cosmological singularity.

\begin{stat}\label{stat_sing_points}
The dynamical system \eqref{dxi/dt}, \eqref{dPhi/dt},  \eqref{dZ/dt}, \eqref{dz/dt}, \eqref{dH/dt_0}, \eqref{Surf_Einst1} has singular points only for special values of the cosmological constant $\Lambda_0$:
\begin{enumerate}
\item при $\displaystyle\Lambda_0=\frac{m^4}{4\alpha}+\frac{\mathfrak{m}^4}{4\beta},\quad (\Lambda=0)$\\
 -- straight line of singular points $M_\xi$ \eqref{M-xi}, corresponding to null values of scalar potentials;
\item при $\Lambda_0=0$ $\biggl(\displaystyle\Lambda=-\frac{m^4}{4\alpha}-\frac{\mathfrak{m}^4}{4\beta}\biggr)$\\
-- 4 singular points in the infinite future $M_{\pm,\pm}$ \eqref{M_pm_pm} with nonzero values of scalar potentials;
\item при $\displaystyle\Lambda_0=\frac{\mathfrak{m}^4}{4\beta}$ $\displaystyle\biggl(\Lambda=-\frac{m^4}{4\alpha}\biggr)$\\
-- 2 singular points in the infinite future $M_{\pm,0}$ \eqref{M_pm_0} with nonzero value of the potential of phantom field;
\item при $\displaystyle\Lambda_0=\frac{m^4}{4\alpha}$ $\displaystyle\biggl(\Lambda=-\frac{\mathfrak{m}^4}{4\beta}\biggr)$\\
-- 2 singular points in the infinite future $M_{0,\pm}$ \eqref{M_0_pm} with nonzero values of the classical field's potential.
\end{enumerate}
Thus, the dynamic system \eqref{dxi/dt}, \eqref{dPhi/dt},  \eqref{dZ/dt}, \eqref{dz/dt}, \eqref{dH/dt_0} in the case of an arbitrary value of the cosmological constant in general has no singular points, for special values of the cosmological constant it can have a straight line of singular points, either 4 or 2 singular points in the infinite future.
\end{stat}

\subsection{Eigenvalues of the Matrix of the Dynamic System for the Model 1}
Let us turn to the study of the nature of the singular points in those cases when these points exist. Calculating the main matrix of the dynamical system \eqref{dxi/dt}, \eqref{dH/dt}, \eqref{dPhi/dt}, \eqref{dZ/dt}, \eqref{dz/dt}
\[A=\left|\left|\frac{\partial X_i}{\partial x_k}\right|\right|\]
in singular points \eqref{Zz=0} and \eqref{H=0}, we find:
\begin{equation}\label{A(M)}
A(M)=\displaystyle\left(\begin{array}{cccccc}
0 & 1 & 0 & 0 &0 & 0 \\
\frac{\partial P_1}{\partial \xi} & 0 & \frac{\partial P_1}{\partial \Phi} & 0 &  \frac{\partial P_1}{\partial \varphi} & 0\\
0 & 0 & 0 & 1 & 0 & 0 \\
\frac{\partial P_2}{\partial \xi} & 0 & \frac{\partial P_2}{\partial \Phi} & 0 & 0 & 0\\
0 & 0 & 0 & 0 & 0 & 1 \\
\frac{\partial P_3}{\partial \xi} & 0 & 0 & 0 & \frac{\partial P_3}{\partial \varphi} & 0 \\
\end{array}\right)_M,
\end{equation}
where $P_1$, $P_2$ and $P_3$ -- are right parts of the equations \eqref{dH/dt},  \eqref{dZ/dt} and \eqref{dz/dt}, correspondingly. It can be seen that in general case the matrix \eqref{A(M)} is not degenerate.

Using the obvious relations
\[\frac{\partial \psi_{(a)}}{\partial \xi}=-\psi_{(a)};\; \frac{\partial \psi_{(a)}}{\partial \Phi_r}=-\frac{\psi_{(a)}}{\Phi_r}, \]
and also the definition of the functions $P_1$ \eqref{dH/dt_0}, $P_2$ \eqref{dZ/dt}, $P_3$ \eqref{dz/dt} and expressions \eqref{F'_12} for derivatives functions $F_1(x)$ and $F_2(x)$, we obtain expressions for the partial derivatives of the functions $P_k$, included in the matrix $A(M)$:
\[\frac{\partial P_1}{\partial \xi}=-\frac{4e^4\Phi^4\psi^3_c}{3\pi\sqrt{1+\psi^2_c}}(3+4\psi^2_c)-
\frac{4\epsilon^4\varphi^4\psi^3_f}{3\pi\sqrt{1+\psi^2_f}}(3+4\psi^2_f); \]
\[\frac{\partial P_1}{\partial \Phi}=-\frac{4\pi^3_c \mathrm{e}^{-3\xi}}{3\pi\sqrt{1+\psi^2_c}};\quad
\frac{\partial P_1}{\partial \varphi}=-\frac{\pi^3_f \mathrm{e}^{-3\xi}}{3\pi\sqrt{1+\psi^2_\epsilon}};  \]
\[\frac{\partial P_2}{\partial \xi}=\mathrm{sgn}(e\Phi)\frac{8e^2\pi^3_c}{\pi\sqrt{1+\psi^2_c}}\mathrm{e}^{-3\xi}; \:
\frac{\partial P_3}{\partial \xi}=\mathrm{sgn}(\epsilon\varphi)\frac{8\epsilon^2\pi^3_f}{\pi\sqrt{1+\psi^2_f}}\mathrm{e}^{-3\xi}; \]
\[ \frac{\partial P_2}{\partial \Phi}=-m^2+3\alpha\Phi^2-\frac{4}{\pi}e^4\Phi^2\biggl(3F_1(\psi_c)-
\frac{2\psi^3_c}{\sqrt{1+\psi^2_c}}\biggr);  \]
\[ \frac{\partial P_3}{\partial \varphi}=\mathfrak{m}^2-3\beta\varphi^2+\frac{4}{\pi}\epsilon^4\varphi^2\biggl(3F_1(\psi_f)-
\frac{2\psi^3_f}{\sqrt{1+\psi^2_f}}\biggr),  \]
where $\mathrm{sgn}(x)$ -- is a sign function
\[\mathrm{sgn}(x)=\left\{\begin{array}{rl}
-1, & x<0\\
0, & x=0 \\
1, & x>0 \\
\end{array}\right. .
\]

Calculating the value of these coefficients of the matrix $A(M)$ at limit values $\xi,\Phi,\varphi$ taking into account the asymptotic formulas \eqref{F,x->0} --
\eqref{F,x->8}, we find:
\begin{eqnarray}\label{asymp_matr}
\left.\frac{\partial P_1}{\partial \xi}\right|_{\xi\to\infty}=0; \left.\frac{\partial P_2}{\partial \xi}\right|_{\xi\to\infty}=0;
\left.\frac{\partial P_3}{\partial \xi}\right|_{\xi\to\infty}=0;\left.\frac{\partial P_1}{\partial \Phi}\right|_{\Phi\to0}=0;
\left.\frac{\partial P_1}{\partial \varphi}\right|_{\varphi\to0}=0;\nonumber\\
 \left.\frac{\partial P_1}{\partial \xi}\right|_{\Phi,\varphi\to0}=-\frac{4}{3\pi}(\pi^4_c+\pi^4_f)\mathrm{e}^{-3\xi};
\left.\frac{\partial P_2}{\partial \Phi}\right|_{\Phi\to0}=-m^2; & \displaystyle \left.\frac{\partial P_3}{\partial \varphi}\right|_{\varphi\to0}=\mathfrak{m}^2;\nonumber\\
\left.\frac{\partial P_2}{\partial \xi}\right|_{\Phi\to0}\simeq \frac{8e^3\pi^2_c}{\pi}\Phi \mathrm{e}^{-3\xi}\to 0;
\left.\frac{\partial P_3}{\partial \xi}\right|_{\varphi\to0}\simeq \frac{8\epsilon^3\pi^2_f}{\pi}\varphi \mathrm{e}^{-3\xi}\to 0;\nonumber\\
\left.\frac{\partial P_1}{\partial \Phi}\right|_{\xi\to\infty}=0; \left.\frac{\partial P_1}{\partial \varphi}\right|_{\xi\to\infty}=0; \left.\frac{\partial P_2}{\partial \Phi}\right|_{\xi\to\infty}=-m^2+3\alpha\Phi^2;
\left.\frac{\partial P_3}{\partial \varphi}\right|_{\xi\to\infty}=\mathfrak{m}^2-3\beta\varphi^2;\nonumber\\
\left.\frac{\partial P_2}{\partial \Phi}\right|_{\Phi\to0}=-m^2-\frac{4e^2\pi^2_c}{\pi};
\left.\frac{\partial P_3}{\partial \varphi}\right|_{\varphi\to0}=\mathfrak{m}^2+\frac{4\epsilon^2\pi^2_f}{\pi}.
\end{eqnarray}

\subsection{The Character of the Straight Line of Singular Points $M_\xi$ ($\Lambda=0$)}

Calculating these coefficients of the matrix $A(M_\xi)$ taking into account limit relations \eqref{asymp_matr}, let us find in point $M_\xi$
\begin{equation}\label{A(M_xi)}
A(M_\xi)=\displaystyle\left(\begin{array}{cccccc}
0 & 1 & 0 & 0 &0 & 0 \\
\frac{\partial P_1}{\partial \xi} & 0 & 0 & 0 & 0 & 0\\
0 & 0 & 0 & 1 & 0 & 0 \\
0 & 0 & \frac{\partial P_2}{\partial \Phi} & 0 & 0 & 0\\
0 & 0 & 0 & 0 & 0 & 1 \\
0 & 0 & 0 & 0 & \frac{\partial P_3}{\partial \varphi} & 0 \\
\end{array}\right)_{M_\xi},
\end{equation}
where
\[\left.\frac{\partial P_1}{\partial \xi}\right|_{M_\xi}=-\frac{4}{3\pi}(\pi^4_c+\pi^4_f)\mathrm{e}^{-3\xi};
\left.\frac{\partial P_2}{\partial \Phi}\right|_{\Phi\to0}=-m^2-\frac{4e^2\pi^2_c}{\pi}; \left.\frac{\partial P_3}{\partial \varphi}\right|_{\varphi\to0}=\mathfrak{m}^2+\frac{4\epsilon^2\pi^2_f}{\pi}. \]
Thus, the matrix of the dynamical system is not degenerate, its determinant is equal to:
\[\mathrm{det}(A(M_\xi))=-\frac{\partial P_1}{\partial \xi}\frac{\partial P_2}{\partial \Phi}\frac{\partial P_3}{\partial \varphi}<0.\]
Since the matrix $A(M_\xi)$ has a block-diagonal structure, its eigenvectors, $k_i$, can be calculated in an elementary way:
\begin{eqnarray}\label{k_xi}
k_{1,2}=\pm i\sqrt{\frac{4}{3\pi}(\pi^4_c+\pi^4_f)\mathrm{e}^{-3\xi}};& \displaystyle k_{3,4}=\pm ш\sqrt{m^2+\frac{4e^2\pi^2_c}{\pi}};
& k_{5,6}=\pm\sqrt{\mathfrak{m}^2+\frac{4\epsilon^2\pi^2_f}{\pi}}.
\end{eqnarray}
\begin{stat}\label{eigen_M_xi}
Thus, the eigenvalues of the matrix of the dynamical system on the line of singular points $M_\xi$ can only be alternating signs, either real or imaginary. In this case, saddle singular points correspond to real alternating-sign eigenvalues, while attracting centers correspond to imaginary sign-alternating ones. This shows that in the two-dimensional directions $\{\xi,H\}$  and $\{\Phi,Z\}$ (classical subspace) the phase trajectories rotate around the singular points $M_\xi$. In this case, the radius of the phase trajectory in the projection $\{\xi,H\}$ decreases with time and tends to zero at $\xi\to+\infty$. Радиус The radius of the phase trajectory in the projection $\{\Phi,Z\}$ remains constant.  In the two-dimensional direction $\{\varphi,z\}$ (phantom subspace) the singular point is a saddle point.
\end{stat}

\subsection{The Character of Singular Points at $\xi\to +\infty$}
Calculating these coefficients of the matrix $A(M)$ at $\xi\to \infty$ taking into account the limit relations \eqref{asymp_matr}, we find for all points  $M_\infty$ \eqref{M_pm_pm} -- \eqref{M_0_pm}:
\begin{equation}\label{A(M_pm_pm)}
A(M_\infty)=\displaystyle\left(\begin{array}{cccccc}
0 & 1 & 0 & 0 &0 & 0 \\
0 & 0 & 0 & 0 & 0 & 0\\
0 & 0 & 0 & 1 & 0 & 0 \\
0 & 0 & \frac{\partial P_2}{\partial \Phi} & 0 & 0 & 0\\
0 & 0 & 0 & 0 & 0 & 1 \\
0 & 0 & 0 & 0 & \frac{\partial P_3}{\partial \varphi} & 0 \\
\end{array}\right)_{M_\infty},
\end{equation}
Thus, at all points of $M_\infty$ the matrix of the dynamical system is degenerate, its rank is 5. In this case, the eigenvalues of the matrix in the two-dimensional direction $\{\xi,H\}$ are equal to zero, the eigenvalues in the two-dimensional directions $\{\Phi,Z\}$ and $\{\varphi,z\}$, generally speaking, are nonzero:
\begin{eqnarray}\label{k_8}
k_{1,2}=0; & k_{3,4}=\pm\sqrt{-m^2+3\alpha\Phi^2}; & k_{5,6}=\pm\sqrt{\mathfrak{m}^2-3\beta\varphi^2}.
\end{eqnarray}
As a result, we find for the singular points \eqref{M_pm_pm} -- \eqref{M_0_pm}:
\begin{eqnarray}\label{k-M_pm_pm}
M_{\infty,\pm,\pm}\; \biggl(\Lambda_0=0,\; \Lambda=-\frac{m^4}{4\alpha}-\frac{\mathfrak{m}^4}{4\beta}\biggr): & k_{1,2}=0;\; k_{3,4}=0;\;k_{5,6}=0;\\
\label{k-M_pm_0}
M_{\infty,\pm,0}\; \biggl(\Lambda_0=\frac{\mathfrak{m}^4}{4\beta};\; \Lambda=-\frac{m^4}{4\alpha}\biggr): & k_{1,2}=0;\; k_{3,4}=0;\; k_{5,6}=\pm \mathfrak{m};\\
\label{k-M_0_pm}
M_{\infty,0,\pm}\; \biggl(\Lambda_0=\frac{m^4}{4\alpha}\biggr)\; \Lambda=-\frac{\mathfrak{m}^4}{4\beta}: & \quad k_{1,2}=0;\; k_{3,4}=\pm im;\; k_{5,6}=0.
\end{eqnarray}

\begin{stat}\label{eigen_M_8}
Thus, in all infinitely distant singular points $\xi\to+\infty$ in two-dimensional direction  $\{\xi,H\}$ the eigenvalues of the dynamic system are zero. In this case, the matrix of the dynamic system has only nonzero eigenvalues in the singular points $M_{\infty,\pm,\pm}$. This means that the solution of dynamic equations tends to the coordinates of this point, i.e., is an asymptotically exact solution:
\begin{equation}\label{xi->8,Mod1,pmpm}
\xi\to+\infty;\; H\to0;\; \Phi\to\pm\frac{m}{\sqrt{\alpha}};\; Z\to 0; \varphi\to \pm\frac{\mathfrak{m}}{\sqrt{\beta}};\; z\to 0\quad (\Lambda_0=0).
\end{equation}
Further, since in the singular points $M_{\pm,0}$ the solution is unstable in the phantom plane, it makes no sense to consider it. In singular points $M_{0,\pm}$ he solution is stable, therefore the cosmological model can reach a phantom final in the infinite future
\begin{equation}\label{xi->8,Mod1,pmpm}
\xi\to+\infty;\; H\to0;\; \Phi\to0;\; Z\to 0; \varphi\to \pm\frac{\mathfrak{m}}{\sqrt{\beta}};\; z\to 0\quad \biggl(\Lambda_0=\frac{m^4}{4\alpha}\biggr)\; \Lambda=-\frac{\mathfrak{m}^4}{4\beta}.
\end{equation}
\end{stat}
\subsection{Asymptotic Behavior of Invariant Cosmological Acceleration in Stable Singular Points}
Let us investigate the asymptotic behavior of the cosmological acceleration \eqref{Omega(H)} in singular points of the infinitely distant future ($\xi\to+\infty$).
Since in the singular points it is $\dot{H}=H^2=0$, when calculating $\Omega(+\infty)$ in the formula \eqref{Omega(H)} it is necessary to calculate the limit at $\xi\to+\infty$. We find the asymptotic expressions in these points from the formulas \eqref{dH/dt_0} -- \eqref{Surf_Einst1}
\begin{eqnarray}\label{H2,xi8}
3\left.H^2\right|_{\xi\to+\infty}\simeq \Lambda+\frac{m^2\Phi^2_0}{2}-\frac{\alpha\Phi^4_0}{4}+\frac{m^2\varphi^2_0}{2}-\frac{\beta\varphi^4_0}{4}+
\frac{8}{3\pi}\mathrm{e}^{-3\xi}(\pi^3_c|e\Phi_0|+\pi^3_f|\epsilon\varphi_0|);\\
\label{dotH,xi8}
\left.\dot{H}\right|_{\xi\to+\infty}\simeq -\frac{4}{3\pi}\mathrm{e}^{-3\xi}(\pi^3_c|e\Phi_0|+\pi^3_f|\epsilon\varphi_0|).
\end{eqnarray}
It is easy to see that both on the line of singular points \eqref{M-xi} and in all singular points \eqref{M_pm_pm} -- \eqref{M_0_pm} the expression for $H^2$ \eqref{H2,xi8}, containing the cosmological constant and Higgs potential, turns to zero. Thus, let us find with a help of \eqref{Omega(H)}, \eqref{H2,xi8} and \eqref{dotH,xi8} in all singular points of the infinite future:
\begin{equation}\label{Omega,xi8|}
\Omega(\xi\to +\infty)=-\frac{1}{2}; \quad
\bigl[ M_\xi(\xi\to+\infty);\; M_{\infty,\pm,\pm},\; M_{\infty,0,\pm} \bigr].
\end{equation}
Therefore, the following statement is true.
\begin{stat}\label{stat_Omega8_Mod1}
All stable singular points in Model 1 correspond to the effective summary nonrelativistic equation of state in the infinite future.
\end{stat}

\section*{The Conclusion}
Let us now sum up and analyze the results of the research. The following results are presented in the paper.\\[6pt]
\Vivod{Based on both the Lagrangian and Hamiltonian formalism, the equations of motion of a relativistic scalar charged particle situated in gravitational and scalar fields, are obtained. The expression for the dynamic mass of a particle that satisfies the principles of additivity of the Lagrange function and charge conjugation, is obtained. It is shown that the indicated fundamental principles require the removal of the restriction on the nonnegativity of the dynamic mass of a scalar charged particle. It is shown that at transition to the sector of negative values of the dynamic mass of a particle, its four-dimensional velocity vector changes its orientation, however, the observable physical quantities: the total mass of a particle, its momentum vector and three-dimensional velocity vector conserve their classical values.}
\Vivod{A self-consistent system of Einstein equations and scalar fields with microscopic singular densities of a system of scalar charged particles is obtained on the basis of the Lagrangian formalism. The expressions for the microscopic energy density of a scalar charge, the microscopic vector of the scalar current density, and the microscopic tensor of energy - momentum of particles are obtained. The laws of conservation of scalar charge and the total microscopic tensor of energy - momentum of a system of self-gravitating scalar charged particles are obtained. Equations of motion of scalar charged particles are obtained from the microscopic Einstein equations. It is shown that the obtained formula for the dynamic mass provides the correct expression for the tensor of energy - momentum of a system of particles. }
\Vivod{A particle distribution function, singular on the mass surface, is determined on the 8-dimensional phase space of particles, which represents the stratification.  With the help of this function and the timelike field of observers, the macroscopic averages of dynamic functions are determined by means of integration over the phase space. In this case, the connection between the proper time of particles and the proper time of observers is carried out by means of the $\delta$-function, and the transition to the 7-dimensional phase space and the nonsingular distribution function on the mass surface is carried out by means of the $\delta$-function of the Hamiltonian. As a result, the macroscopic densities are correctly and invariantly determined: the scalar charge density, the current vector, and the energy-momentum tensor of the particles. The self-consistent macroscopic equations of gravitational and scalar fields and laws of conservation of charge and total energy - momentum of the system are obtained. Thus, on the basis of microscopic, both Lagrangian and Hamiltonian particle dynamics, a macroscopic description of the statistical system of scalar charged particles is substantiated.
}
\Vivod{The local thermodynamic equilibrium of a statistical system of scalar charged particles is investigated and the formulas are obtained for the equilibrium values of macroscopic scalars - the energy density of particles, pressure, particle number density, and scalar charge density. The transformation properties of these scalars are studied in relation to the transformations of charge conjugation and reflection of scalar fields. It is shown that, due to the invariant properties of these transformations, the signs of the values of the scalar fields of the statistical system do not depend on the signs of the scalar charges of the particles, but are determined only by the initial conditions.  On the basis of the transport equations, which are a strict consequence of the relativistic kinetic equations, the material equations are obtained - the equations of hydrodynamics and the conservation of the current of scalar charges. The expressions in the elementary functions of all macroscopic scalars for a completely degenerate system of fermions are found.}
\Vivod{Based on the obtained macroscopic equations for a degenerate system of scalar charged fermions, a cosmological model of a spatially flat Universe is investigated, for which a complete system of $n+N+2$ ordinary differential equations is formulated ($n$ -- is the number of types of scalar charged particles, $N$ -- is the number of scalar fields. In addition, there is one integral condition for the solution of the system. An exact solution of these $n$ material equations is found, which expresses the Fermi momentum conservation law. As a result, for the case of an asymmetric scalar doublet $N=2$ a normal autonomous system of 6 ordinary differential equations with respect to 6 dynamical functions is obtained. Two simple models of the Fermi system are proposed: in the first model there are 2 types of fermions, one of which carries a classical, the other - a phantom charge, in the second model there are only one-sort fermions charged simultaneously with two charges.)
}
\Vivod{A complete system of dynamic equations for model 1 is obtained and its qualitative analysis is carried out, the results of which can be briefly described as follows. For arbitrary values ??of the fundamental constants (the structure constants of the Higgs potentials and the cosmological constant), the dynamical system has no singular points. With special ratios between fundamental constants, the model has singular points in the infinite future or straight lines of singular points. This property significantly distinguishes the considered cosmological models from simpler models with a vacuum scalar doublet (see \cite{Ignat21_TMP}).
}
\Vivod{It is shown that all stable singular points in the infinite future $\to 0, \ a \to \infty, \ H  \to 0 $ correspond to the total effective nonrelativistic equation of state $ \Omega = - \frac{1}{2} \Rightarrow w = 0 $. Thus, taking into account the interaction of scalar fields with degenerate fermions can lead to deceleration of the cosmological acceleration in the future history of the Universe. However, it must be emphasized that this conclusion is valid only with a special choice of the fundamental constants of the model.}

Since it is difficult to count on the fact that the real Universe corresponds to the above specific sets of fundamental constants, it is unlikely, and the real cosmological model will have stable singular points. In this regard, one can hope that the consideration of models with scalar charged particles will remove the conclusion $\Omega (t \to + \infty) \to 1$, which is unacceptable for observational cosmology. Therefore, it is necessary to carry out numerical integration of the presented mathematical models in a wide range of fundamental constants. We intend to publish the results of such a study in the near future.

\subsection*{Funding}
The work was carried out at the expense of a subsidy allocated within the framework of state support of the Kazan (Volga Region) Federal University with the aim to increase its competitiveness among the world's leading research and educational centers.

The work is performed according to the Russian Government Program  of Competitive  Growth  of  Kazan  Federal  University.


\end{document}